\documentclass[11pt,preprint]{aastex}

\newcommand{\suzaku}{{\it Suzaku}}
\newcommand{\swift}{{\it Swift}}

\newcommand {\be} {\begin{equation}}
\newcommand {\ee} {\end{equation}}

\slugcomment{e-mail: kataoka@hp.phys.titech.ac.jp}
\slugcomment{Accepted by ApJ}

\shorttitle{Multiwavelength Observations of PKS~1510$-$089}
\shortauthors{Kataoka et al.}

\begin{document}

\title{Multiwavelength Observations of the Powerful $\gamma$-ray 
Quasar\\ 
PKS~1510$-$089: Clues on the Jet Composition}

\author{J.~Kataoka\altaffilmark{1}, G.~Madejski\altaffilmark{2},
M.~Sikora\altaffilmark{3}, P.~Roming\altaffilmark{4}, 
M.~M.~Chester\altaffilmark{4}, D.~Grupe\altaffilmark{4},\\ 
Y.~Tsubuku\altaffilmark{1}, 
R.~Sato\altaffilmark{1}, N.~Kawai\altaffilmark{1},
G.~Tosti\altaffilmark{5}, D.~Impiombato\altaffilmark{5},\\
Y.~Y.~Kovalev\altaffilmark{6,7}, Y.~A.~Kovalev\altaffilmark{7}, 
P.~G.~Edwards\altaffilmark{8}, 
S.~J.~Wagner\altaffilmark{9},\\
R.~Moderski\altaffilmark{3}, 
\L.~Stawarz\altaffilmark{2,10},
T.~Takahashi\altaffilmark{11},  and S.~Watanabe\altaffilmark{11}, 
}

\altaffiltext{1}{Department of Physics, Tokyo Institute of Technology, 
Meguro, Tokyo, 152-8551, Japan}
\altaffiltext{2}{Stanford Linear Accelerator Center and 
Kavli Institute for Particle Astrophysics and
Cosmology, Stanford University, Stanford, CA 94305, USA}
\altaffiltext{3}{Nicolaus Copernicus Astronomical Center, Bartycka 18,
00-716, Warsaw, Poland}
\altaffiltext{4}{Department of Astronomy and Astrophysics, Pennsylvania
State University, PA, 16802, USA}
\altaffiltext{5}{Physics Department and Astronomical Observatory,
University of Perugia, Perugia, Italy}
\altaffiltext{6}{Max-Planck-Institut f\"ur Radioastronomie,
Auf dem H\"ugel 69, 53121 Bonn, Germany}
\altaffiltext{7}{Astro Space Center of the Lebedev Physical Institute, 
Profsoyuznaya 84/32, Moscow, 117997, Russia}
\altaffiltext{8}{Australia Telescope National Facility, CSIRO, Locked
Bag 194, Narrabri NSW 2390, Australia}
\altaffiltext{9}{Landessternwarte Heidelberg, K\H{o}nigstuhl, D-69117
Heidelberg, Germany}
\altaffiltext{10}{Astronomical Observatory of the Jagiellonian University, 
ul. Orla 171, 30-244 Krak\'ow, Poland}
\altaffiltext{11}{Department of High Energy Astrophysics, ISAS, JAXA, Kanagawa, 229-8510, JAPAN}

\begin{abstract}
We present the results from a multiwavelength campaign 
conducted in August 2006 
of the powerful $\gamma$-ray quasar PKS~1510$-$089 ($z$ = 0.361). 
This campaign commenced with a deep {\suzaku} observation 
lasting three days for a total exposure time of 120\,ks, 
and continued with {\swift} monitoring over 18 days. Besides {\swift}
observations, which sampled the optical/UV flux in all 6 UVOT filters as well as 
the X-ray spectrum in the 0.3$-$10\,keV energy range, 
the campaign included ground-based optical 
and radio data, and yielded a quasi-simultaneous broad-band spectral energy 
distribution from 10$^9$\,Hz to 10$^{19}$\,Hz. 
Thanks to its low instrumental background, the {\suzaku} observation provided a 
high S/N X-ray spectrum, which is  
well represented by an extremely hard power-law 
with photon index $\Gamma$\,$\simeq$\,1.2, augmented by a soft component 
apparent below 1\,keV, which is well described by a black-body model 
with temperature $kT$\,$\simeq$\,0.2\,keV.
Monitoring by {\suzaku} revealed temporal variability which is different 
between the low and high energy bands, again suggesting the presence of a 
second, variable component in addition to the primary power-law emission. 
We model the broadband spectrum of PKS~1510$-$089
assuming that the high energy spectral
component results from Comptonization of infrared radiation produced
by hot dust located in the surrounding molecular torus. 
In the adopted internal shock scenario, the derived model parameters imply
that the power of the jet is dominated by 
protons but with a number of electrons/positrons exceeding a number of 
protons by a factor $\sim 10$.
We also find that inhomogeneities responsible for the shock formation, 
prior to the collision may 
produce bulk-Compton radiation which can explain  
the observed soft X-ray excess and possible excess at $\sim 18$ keV. 
We note, however, that the bulk-Compton interpretation is not unique, and the 
observed soft excess could arise as well via some other processes discussed 
briefly in the text.  
\end{abstract}


\section{Introduction}

Powerful, highly-collimated outflows called jets are commonly observed in a 
wide variety of astronomical sources;  for  active galactic nuclei 
(AGNs; e.g., Urry \& Padovani 1995), $\gamma$-ray bursts (e.g., Piran 2000), 
and binaries containing compact stars (e.g., Mirabel \& Rodriguez 1999) we 
have excellent evidence that these outflows move with relativistic speeds.  
It has been a long-standing mystery, however, where and how the 
relativistic jets are formed, and what is their composition.  
From a theoretical standpoint, the relativistic AGN jets considered here 
can be launched as outflows dominated by Poynting flux generated in the 
force-free magnetospheres of black holes, or as hydromagnetic winds driven 
centrifugally from accretion discs (see review by Lovelace, 
Ustyugova, \& Koldova 1999). In either case, strong magnetic 
fields are involved in driving the outflows, although many (if not most) 
observations indicate that eventually, {\sl particles} 
carry the bulk of the jet's energy 
(Wardle et al.\ 1998; Sikora \& Madejski 2000; 
Hirotani 2005; Sikora et al.\ 2005). This apparent discrepancy, 
however, can be resolved if the jets are indeed initially dominated by 
the Poynting flux but are efficiently converted into 
matter-dominated form at some later stage, most likely 
prior to the so-called ``blazar zone'' (see Sikora et al.\ 2005 and 
references therein). Such a ``blazar zone,'' the region 
where the bulk of the observed nonthermal radiation is produced, 
is most likely located at $r$ $\simeq$ 10$^3$$-$10$^4$ $r_g$, 
where $r_g$ = $GM$/$c^2$ is the gravitational radius (Spada et al.\ 2000; 
Kataoka et al.\ 2001; Tanihata et al.\ 2003).

Here we adopt a scenario where a jet is launched near a rapidly rotating
black hole, presumably at the innermost portions of the accretion disk
(see, e.g., Koide et al.\ 1999). Such a jet, initially consisting of 
protons and electrons, is accelerated by large scale magnetic field 
stresses and within 100 $r_g$ can be loaded  by 
electron/positron ($e^{-}e^{+}$) pairs via 
interactions with the coronal soft $\gamma$-ray photons 
(note that such photons are directly seen in the spectra of Seyfert 
galaxies; see, e.g., Zdziarski, Poutanen \& Johnson 2000). 
Hence, it is possible that relativistic jets in quasars (beyond
the jet formation zone) may well contain more electron/positron 
pairs than protons, but are still dynamically 
dominated by cold protons (Sikora et al.\ 1997; Sikora \& Madejski 2000). 

Observations with the EGRET instrument on board the {\it Compton 
Gamma-Ray Observatory} in the $\gamma$-ray band have opened a new window 
for studying AGN jets, and revealed that many radio-bright and variable 
AGN are also the brightest extragalactic MeV$-$GeV $\gamma$-ray emitters 
(see, e.g., Hartman et al.\ 1999).  The properties of the $\gamma$-ray 
emission in those objects --- often termed ``blazars'' --- supported earlier 
inferences based on radio and optical data, and independently indicated  
significant Doppler boosting, implying the origin of broad-band emission 
in a compact, relativistic jet pointing close to our line of sight. 
Generally, the overall spectra of blazar sources (plotted in 
the log$(\nu)$-log$(\nu F_{\nu})$ plane, where $F_{\nu}$ is 
the observed spectral flux energy density) have two pronounced continuum 
components: one peaking between IR and X-rays and 
the other in the $\gamma$-ray regime (see, e.g., Kubo et al.\ 1998; Ghisellini 
et al.\ 1998). The lower energy component is believed to be produced 
by the synchrotron radiation of relativistic electrons accelerated 
within the outflow, while inverse Compton (IC) emission by the same 
electrons is most likely responsible for the formation of the high energy 
$\gamma$-ray component.  It is widely believed, in addition, 
that the IC emission from flat spectrum radio 
quasars (FSRQ) is dominated by the scattering of soft photons external 
to the jet (external Compton process, ERC), which are produced by 
the accretion disk, either directly or indirectly via scattering/reprocessing 
in the broad line region (BLR) or dusty torus (see, e.g., Dermer \& 
Schlickeiser 1993; Sikora, Begelman, \& Rees 1994). Other sources 
of seed photons can also contribute to the observed IC component, 
in particular the synchrotron photons themselves via the synchrotron 
self-Compton process (SSC) which under certain conditions can even dominate 
the observed high energy radiation (Sokolov \& Marscher 2005).  
Detailed modelling of  broad-band blazar emission can provide 
information about the location of the dissipative regions in blazars,
the energy distribution of relativistic electrons/positrons,
the magnetic field intensity, and the jet power.

A probe of the low energy electron/positron content in blazars was
proposed by Begelman \& Sikora (1987), and extensively studied in
the literature (Sikora \& Madejski 2000; Moderski et al.\ 2004; Celotti et al.
2007)\footnote{See also Georganopoulos et al.\ (2005) and Uchiyama et al.\ 
(2005) for applications of the bulk-Compton constraints on the parameters
of AGN jets on large-scales.}. The $\gamma$-ray 
emission is produced by electrons/positrons accelerated {\it in situ}, and thus 
before reaching the blazar dissipative site the 
electrons/positrons are expected to be cold. If they are transported
by a jet with a bulk Lorentz factor 
$\Gamma_{\rm jet} \ge 10$, 
they upscatter external UV photons up to X-ray energies
and produce a relatively narrow feature expected to be located 
in the soft/mid X-ray band, with 
the flux level reflecting the amount of cold electrons and 
the jet velocity.  Unfortunately, such an additional 
bulk-Compton (BC) spectral component is difficult to observe because 
of the presence of strong non-thermal blazar emission, which dilutes 
any other radiative signatures of the active nucleus. In this context, 
FSRQs may constitute a possible exception, since their 
non-thermal X-ray emission is relatively weak when compared to other 
types of blazar sources.

PKS~1510$-$089 is a nearby ($z$ = 0.361) highly polarized quasar (HPQ) 
detected in the MeV$-$GeV band by EGRET. It is a highly
superluminal jet source, with apparent velocities of $v_{\rm app}$ 
$\gtrsim$ 10 $c$ observed in multi-epoch VLBA observations 
(Homan et al.\ 2001; Wardle et al.\ 2005; Jorstad et al.\ 2005). 
Its broad-band 
spectrum is representative of other FSRQs. In particular, 
its radiative output is dominated by the $\gamma$-ray inverse-Compton 
component, while its synchrotron emission peaks around IR frequencies 
below the pronounced UV bump, which is in turn presumably due to the thermal emission 
from the accretion disk (Malkan \& Moore 1986; Pian \& Treves 1993). 
PKS~1510$-$089 has been extensively studied by X-ray satellites,
especially {\it ROSAT} (Siebert et al.\ 1996), {\it ASCA} (Singh, 
Shrader, \& George 1997), and {\it Chandra} (Gambill et al.\ 2003).  
The observed X-ray spectrum was very flat in the 2$-$10\,keV band (photon
index\,$\Gamma$\,$\simeq$ 1.3), but steepened (to $\Gamma$\,$\simeq$\,1.9)
in the {\it ROSAT} bandpass (below 2\,keV). Recent observations by 
{\it BeppoSAX} (Tavecchio et al.\ 2000) confirmed the presence of a 
soft X-ray excess below 1\,keV. All these findings suggest that PKS~1510$-$089 
may be among the best candidates for detecting the BC bump. 

In this paper, we present a detailed analysis of 120\,ks observations of 
PKS~1510$-$089 with {\suzaku} in August 2006 as a part of the AO-1 program, 
in addition to a long {\swift} XRT/UVOT monitoring campaign followed-up by 
ground-based optical and radio telescopes. 
Thanks to the good photon statistics and the low background of the 
{\suzaku}/{\swift} 
data, we successfully obtained the highest quality data on PKS~1510$-$089 
ever reported, over ten decades in frequency, between 10$^9$\,Hz and
10$^{19}$\,Hz. 
The observation and analysis methods are described in $\S$2. 
Detailed spectral studies and temporal analysis 
are presented in $\S$3. Based on these new findings, in $\S$4 we discuss the 
nature of the observed spectral features. Finally, our main 
conclusions are given in $\S$5. Throughout this paper we adopt a luminosity 
distance of $d_{\rm L}$ = 1919\,Mpc for PKS~1510$-$089 ($z$ = 0.361), 
derived for a modern cosmology with $\Omega_{\rm m}$ = 0.27,  
$\Omega_{\rm \Lambda}$ = 0.73 and $H_0$ = 71\,km\,s$^{-1}$\,Mpc$^{-1}$.

\section{Observations and Data Reduction}

\subsection{\suzaku}

PKS~1510$-$089 was observed with {\suzaku} (Mitsuda et al.\ 2007) 
in August 2006 over approximately three days. 
Table~1 summarizes the start and end times, and the 
exposures of the {\suzaku} observation (sequence number 701094010). 
{\suzaku} carries four sets of X-ray telescopes (Serlemitsos et al.\ 2007) 
each with a focal-plane X-ray CCD camera 
(XIS, X-ray Imaging Spectrometer; Koyama et al.\ 2007) that is sensitive 
in the energy range of 0.3$-$12\,keV, together with a non-imaging 
Hard X-ray Detector (HXD;  Takahashi et al.\ 2007; Kokubun et al.\ 2007), 
which covers the 10$-$600\,keV energy band with Si PIN photo-diodes and GSO 
scintillation detectors. Three of the XIS (XIS 0, 2, 3) detectors have 
front-illuminated (FI) CCDs, while the XIS 1 utilizes a back-illuminated (BI) 
CCD. The merit of the BI CCD is its improved sensitivity in the soft X-ray
energy band below 1\,keV. PKS~1510$-$089 was focused on the nominal
center position of the XIS detectors. 

\subsubsection{XIS Data Reduction}

For the XIS, we analyzed the screened data, reduced via {\suzaku} 
software version 1.2. The reduction followed the prescriptions 
described in `The Suzaku Data Reduction Guide' (also known as $the$ $ABC$ 
$guide$) provided by the {\suzaku} guest observer facility at the 
NASA/GSFC.\footnote{http://suzaku.gsfc.nasa.gov/docs/suzaku/analysis/abc. 
See also seven steps to the {\suzaku} data analysis at 
http://www.astro.isas.jaxa.jp/suzaku/analysis.} 
The screening was based on the following 
criteria: (1) only ASCA-grade 0, 2, 3, 4, 6 events are accumulated, 
while hot and flickering pixels were removed from the XIS image 
using the \textsc{cleansis} script, (2) the time interval after 
the passage of South Atlantic Anomaly (T\_SAA\_HXD) is greater than 500\,s, 
(3) the object is at least 5$^\circ$ and 20$^\circ$ above the rim of the 
Earth (ELV) during night and day, respectively.  
In addition, we also select the data with a cutoff 
rigidity (COR) larger than 6\,GV. After this screening, 
the net exposure for good time intervals is 119.2\,ks.  

The XIS events were extracted from a circular region with a 
radius of $4.3'$ centered on the source peak, whereas the background 
was accumulated in an annulus with inner and outer radii of 
$4.9'$ and $6.3'$, respectively.  We carefully checked that the use of 
different source and background regions did not affect the analysis 
results presented in the next sections, within 1\,$\sigma$ uncertainties.
The response (RMF) and auxiliary (ARF) files are produced using the 
analysis tools \textsc{xisrmfgen} and \textsc{xissimarfgen} developed 
by the {\suzaku} team, which are included in the software package 
HEAsoft version 6.12. We also checked whether our spectral
fitting results (see $\S$3) were consistent with what has been 
obtained using the `standard' RMF and 
ARF files, provided for science working group members, 
for a point source placed on the nominal CCD 
position (\textsc{ae\_xi\{0,1,2,3\}\_20060213.rmf} and
\textsc{ae\_xi\{0,1,2,3\}\_xisnom6\_20060615.arf}). This was done 
after correcting for the degradation of the XIS 
response using the tool \textsc{xisscontamicalc}.  

\subsubsection{HXD/PIN Data Reduction}

The source spectrum and the light curves were extracted from the 
cleaned HXD/PIN event files (version 1.2). 
The HXD/PIN data are processed with basically the same screening 
criteria as those for the XIS, except that ELV\,$\ge$\,5$^\circ$ through 
night and day, T\_SAA\,$\ge$\,500\,s, and COR\,$\ge$\,8\,GV. 
The HXD/PIN instrumental background spectra were generated from a 
time dependent model provided by the HXD instrument team for each 
observation (see Kokubun et al.\ 2007; Fukazawa et al.\ 2006 for 
more details; also see Kataoka et al.\ 2007 concerning the robustness of 
background subtraction of the HXD/PIN using the most recent response and
background models). Both the source and background spectra were made 
with identical good time intervals (GTIs) and the exposure was 
corrected for detector deadtime of 6.3$\%$. We used the response files 
version \textsc{ae\_hxd\_pinxinom\_20060814.rsp}, provided by 
the HXD instrumental team.

\begin{figure}
\begin{center}
\includegraphics[angle=0,scale=.70]{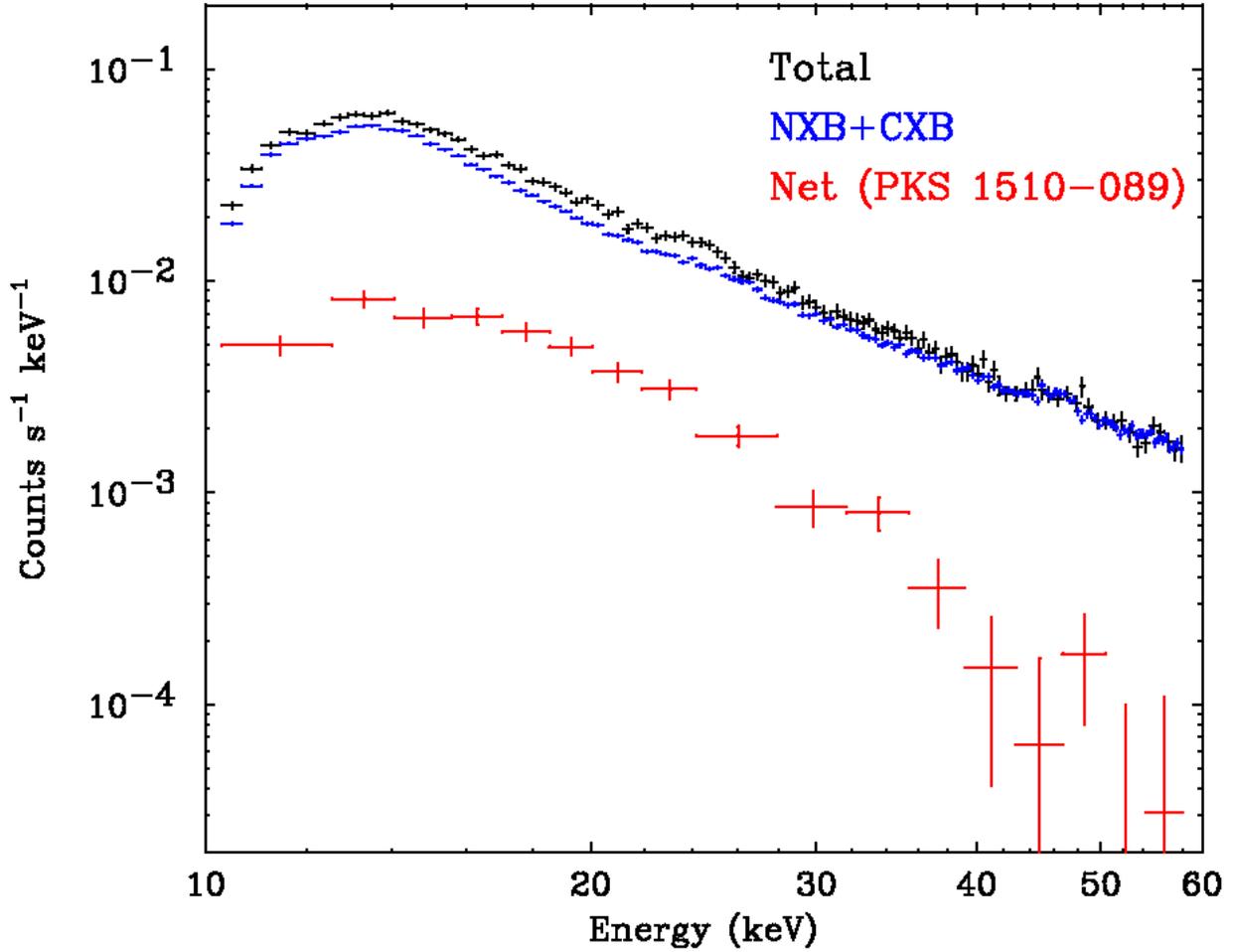}
\caption{The combined HXD/PIN spectra for the {\suzaku} observation of PKS~1510$-$089 
 over the whole HXD/PIN energy bandpass (10$-$60\,keV). $Black$ (upper):
 source plus background spectrum, $blue$ (middle): 
sum of the non--X-ray background
 and CXB, and $red$ (bottom): for the net source spectrum.}\label{fig:pinspec}
\end{center}
\end{figure}

The time averaged HXD/PIN spectrum thus obtained is shown in 
Figure~\ref{fig:pinspec}, plotted over the energy range 10--60\,keV. 
HXD/PIN data below 12\,keV have been ignored to avoid noise contamination 
near the lower threshold of the PIN diode. Also the data above 50\,keV 
are discarded, as a detailed study of noise and background systematics 
is on-going above this energy. Figure \ref{fig:pinspec} shows the total 
(PKS~1510$-$089 + observed background) spectrum, where the background 
includes both the instrumental (non X-ray) background and the 
contribution from the cosmic X-ray background (CXB) 
(Gruber et al.\ 1999; see also Frontera et al.\ 2007 for updated 
{\it BeppoSAX} results).  
Here the form of the CXB was taken as 
9.0$\times$10$^{-9}$(E/3\,keV)$^{-0.29}$\,exp($-$E/40\,keV)\,erg\,cm$^{-2}$\,s$^{-1}$\,sr$^{-1}$\,keV$^{-1}$ 
and the observed spectrum was simulated assuming the PIN detector response 
to isotropic diffuse emission. When normalized to the field of 
view of the HXD/PIN instrument the effective flux of the CXB component 
is expected to be 9.0$\times$10$^{-12}$\,erg\,cm$^{-2}$\,s$^{-1}$ 
in the 12$-$50\,keV band, which is about $\sim$20$\%$ of the PKS~1510$-$089 
flux in the same energy bandpass. 

Assuming that the spectral shape determined by the HXD/PIN (below 
50\,keV) would extend to 100\,keV, the PKS~1510$-$089 flux would be 
$\sim$ 3.5$\times$10$^{-11}$\,erg\,cm$^2$\,s$^{-1}$ (50$-$100\,keV)
and could therefore ultimately be detected by the HXD/GSO detector. However, 
this is only a few percent of the GSO detector background and the 
study of this level of background systematics is still on-going by 
the HXD instrument team. Therefore, in this paper, we do not include 
the GSO data in the subsequent spectral fits. We also note that after 
2006 May 24, bias voltages for 16 out of 64 PIN diodes (in the W0 unit) 
were reduced from 500 to 400\,V to suppress the rapid increase of 
noise events caused by in-orbit radiation
damage\footnote{http://www.astro.isas.jaxa.jp/suzaku/analysis/hxd/hxdresp}.
It is thus recommended that a careful comparison of 
the analysis results by including/excluding W0 unit be made. 
In this paper, we used all HXD/PIN sensors 
(including W0) because no differences were found between both analyses.

\subsection{\swift}

PKS~1510$-$089 was observed with {\swift} (Gehrels et al.\ 2004) 
10 times in August 2006, as a Target of Opportunity (ToO) 
with a total duration of 24.3\,ks.
Table~1 summarizes the start and end times, 
and the exposure time of each observation. Note the {\swift}
observations cover more than 18 days from August 4th to 22nd, and
the first two observations are well within the range of the {\suzaku} 
observation. {\swift} carries 
three sets of instruments, the Burst Alert
Telescope BAT (15$-$150\,keV; Barthelmy et al.\ 2005), the X-ray
Telescope XRT (0.3$-$10\,keV; Burrows et al.\ 2005) and the
Ultra-Violet/Optical Telescope UVOT (170$-$650\,nm; Roming et al.\ 2005).
Hereafter we only analyzed the XRT and the UVOT data because
the source was not detected in the BAT exposures.

\subsubsection{XRT Data Reduction}

The {\swift} XRT data were all taken in Photon Counting mode (PC mode; Hill
et al.\ 2004). The data were reduced by the XRT data analysis task 
\textsc{xrtpipeline} version 0.10.4, which is part of the HEASARC 
software package 6.1. Source photons were selected in a
circle with a radius of $47''$ and background photons in a nearby
source-free circle with a radius of $189''$.  Photons were selected 
from the event file by XSELECT version 2.4. Spectra and light curves 
were corrected for losses due to dead columns (Abbey et al.\ 2006). 
Photons for the spectral analysis with grades 0$-$12 were selected 
and rebinned with \textsc{grppha} version 3.0.0 having at
least 20 photons per bin. The auxiliary response file was created by the
XRT task \textsc{xrtmkarf} and the standard response file \textsc{swxpc0to12\_20010101v008.rmf}. 
All spectra were analyzed in the 0.3$-$10.0\,keV band using XSPEC version 12.3.0
(Arnaud 1996). Due to the low count rate, no correction for pileup was applied.

\subsubsection{UVOT Data Reduction}

Data from the UVOT and XRT were obtained simultaneously. The UVOT
observing mode used one exposure in each of six optical and
ultraviolet filters (in order; uvw2, v, uvm2, uvw1, u, and b) per {\swift}
pointing (typically 5--15 minutes long).  Event data were preserved for
uvm2 and uvw2; for the other filters the data were converted to images
on-board.  Exposures were processed at the {\swift} Data Center.  For this
analysis, Level~1 event data and Level~2 sky-corrected image data were
used.  Photometry was performed using the HEASARC FTOOLs \textsc{uvotevtlc} for
the event data (uvm2 and uvw2) and \textsc{uvotsource} for the image data (v,
b, u and uvw1), and CALDB Version 2006-11-16. Since the source was
relatively bright, the source aperture sizes were chosen to correspond
to those used to determine the UVOT zero-points: 6$''$ for the
optical and 12$''$ for the ultraviolet filters. Therefore, no
aperture correction was required. A 25$''$ background region was
extracted from a blank area of the sky offset from the source.  Several
blank regions were tried; the choice of background region made only very
minor differences to the final results. All event and image data were 
corrected for coincidence loss. All event data for a given orbit 
were binned together. A comparison of the event photometry to the 
ground-processed uvm2 and uvw2 images were made; the results were 
entirely consistent. 

\subsection{Ground-based observations}

\subsubsection{Optical}

The photometric Optical/IR observations were carried out with two
instruments: the Newtonian f/5, 0.4\,m, Automatic Imaging Telescope (AIT)
of the Perugia University Observatory, and the the Rapid Eye Mount (REM, see
Zerbi et al.\ 2004) a robotic telescope located at the ESO Cerro La Silla
observatory in Chile\footnote{http://www.rem.inaf.it}. 
The AIT is based on an equatorially mounted 0.4-m
Newtonian reflector having a 0.15-m reflector solidly joined to it. 
The AIT is a robotic telescope equipped with a 192$\times$165 pixels CCD
array, thermoelectrically cooled with Peltier elements and with
Johnson-Cousins BVRI filters utilized for photometry (Tosti et al.\ 1996).

The REM telescope has a Ritchey-Chretien
configuration with a 60\,cm f/2.2 primary and an overall f/8 focal ratio
in a fast moving alt-azimuth mount providing two stable Nasmyth focal
stations. At one of the two foci the telescope simultaneously feeds, by
means of a dichroic, two cameras: REMIR for the NIR (see Conconi et al.\ 2004), 
and ROSS (see Tosti et al.\ 2004) for the optical.
Both the cameras have a field of view of $10''$\,$\times$\,$10''$ and imaging
capabilities with the usual NIR (z', J, H and K) and Johnson-Cousins VRI
filters. All raw optical CCD frames obtained with the AIT and REM Telescopes,
were corrected for dark, bias and flat field. Instrumental magnitudes
were obtained via aperture photometry using \textsc{daophot} (Stetson 1988) and
\textsc{Sextractor} (Bertin \& Arnouts 1996).
Calibration of the source magnitude was obtained by differential
photometry with respect to the comparison star sequence reported by
Villata et al.\ (1997) and  Raiteri et al.\ (1998).

Also a few optical observations were carried out using a 70\,cm
telescope of the Landessternwarte in Heidelberg, Germany.
Unfortunately most of the observations were not successful due to
bad weather, and we only got data on August 1,
just before the {\suzaku} observation started. We obtained B, R, I
photometric measurements (2 points each, both consistent with each other
within the relative photometric accuracies of 0.01 mag),
suggesting no variations on a timescale of an hour within this night.
The observation log and resultant magnitudes of PKS~1510$-$089 during 
2006 August observations are summarized in Table~2. 

\subsubsection{Radio}

The 1--22~GHz instantaneous radio spectrum of PKS~1510$-$089 was monitored
with the 600-meter ring radio telescope RATAN-600 (Korolkov \& 
Pariiskii 1979) of the Special Astrophysical Observatory, 
Russian Academy of Sciences, on
2006 August 9, 10, 11, 22, and 23. Observations were made at the 
Northern sector of the telescope with the secondary reflector 
of cabin No.~1. The continuum spectrum was measured
at six different frequencies --- 1, 2.3, 4.8, 7.7, 11, and 22\,GHz ---
within two minutes in a transit mode. Details on the
method of observation, data processing, and calibration are described in
Kovalev et al.\ (1999). Since no significant time variations were 
found during these observations, the averaged data of the 
five independent spectral measurements are provided in Table~3.

PKS~1510$-$089 was also observed with the Australia Telescope Compact 
Array (ATCA). The ATCA consists of six 22\,m 
diameter paraboloidal antennas (Frater, Brooks, \& Whiteoak 1992) with five
antennas on a 3\,km east-west track which has a 214\,m north-south spur,
and a sixth antenna fixed 3\,km to the west of the east-west track.
Snapshot observations were made with the ATCA on two dates.
The first observations were made on 2006 July 11 at 
1.4, 2.3, 4.8 and 8.6\,GHz
while the array was in 750B configuration, with a maximum baseline of 4.5\,km.
The second observations were made on August 4 in H168 configuration using 
the inner five telescopes, 
with a maximum baseline of 192\,m, at frequencies of 
18.5 and 19.5\,GHz in addition to the same 
four frequencies as the first epoch. PKS~1934$-$638 was used as 
the primary flux density and bandpass calibrator at both epochs.
Data were reduced with \textsc{miriad} and the flux densities calculated
under the assumption the source is point-like using the task
\textsc{uvflux} (Sault et al.\ 1995). 
The flux densities at both the first and the second epochs 
are summarized in Table~3. The errors in the flux densities 
include the intrinsic scatter in the data and a conservative allowance 
of 5$\%$ for systematic variation in the flux density scale 
(Tingay et al.\ 2003).

\section{Results}

\subsection{\suzaku}

\begin{figure}
\begin{center}
\includegraphics[angle=0,scale=.70]{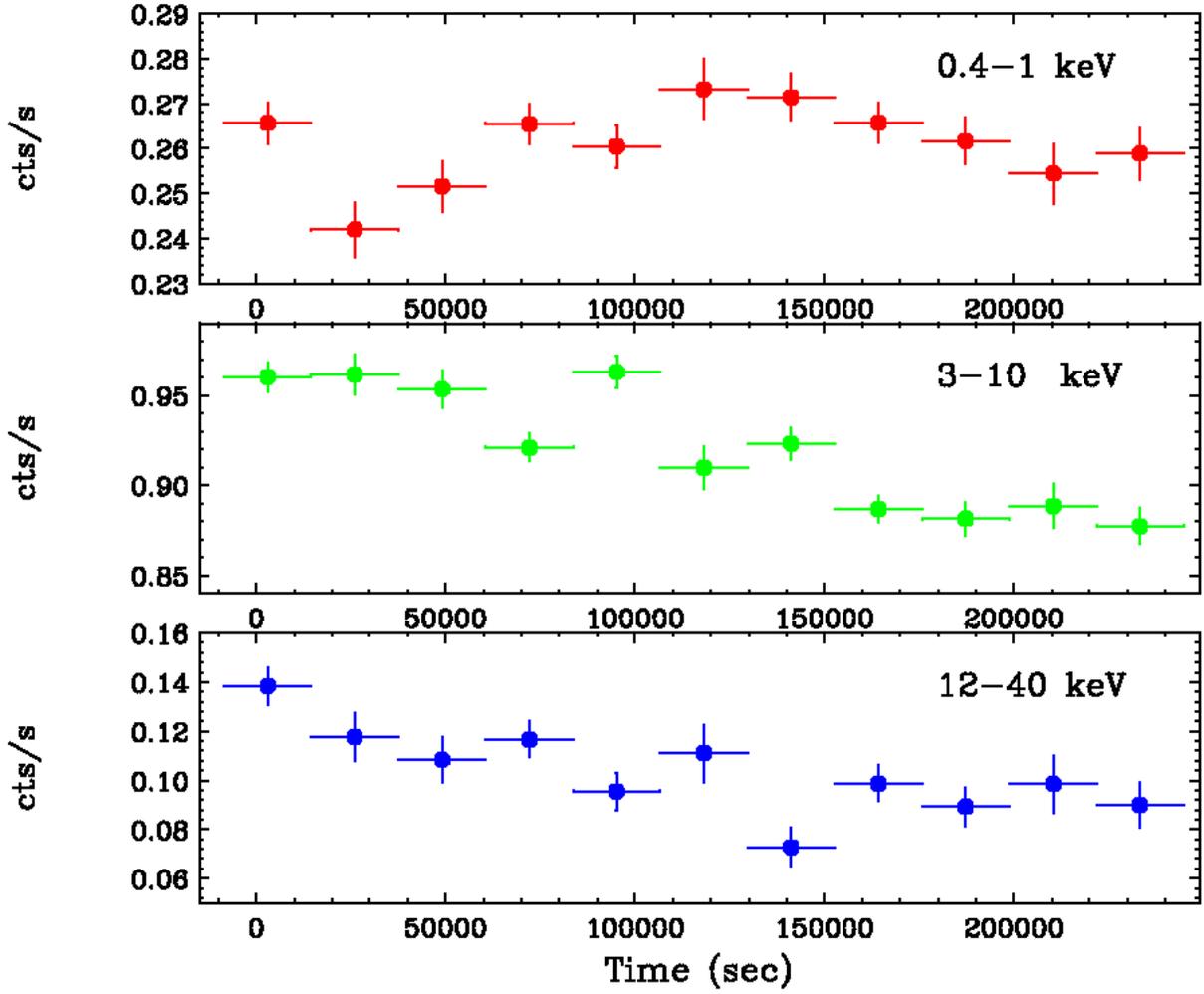}
\caption{The overall variability of PKS~1510$-$089 observed with 
{\suzaku} in 2006 August. $upper$ $panel$: 0.4$-$1\,keV (XIS 0$-$3 summed), 
$middle$ $panel$: 3$-$10\,keV(XIS 0$-$3 summed), and $lower$ $panel$: 
12$-$40\,keV (HXD/PIN W0$-$3 summed).}\label{fig:suzakulc}
\end{center}
\end{figure}

During the {\suzaku} observation (August 2$-$5; Table~1), 
PKS~1510$-$089 was in a relatively bright state with the average net 
count rates of 4 XISs, measured in the 0.4$-$10\,keV range, 
of 2.388 $\pm$ 0.003\,cts\,s$^{-1}$. 
For the PIN detector, the net average source count rate in the 
12$-$40\,keV band was 0.105 $\pm$ 0.003\,cts\,s$^{-1}$, compared to the PIN 
background (non X-ray background) rate of 0.425\,cts\,s$^{-1}$.
Figure~\ref{fig:suzakulc} shows count rate variations during the 
{\suzaku} observation. The light curves of the 4 XISs and the HXD/PIN 
detectors are shown separately in different energy bands; 
0.4$-$1\,keV ({\it upper}; XIS), 3$-$10\,keV ({\it middle}; XIS) and 
12$-$40\,keV ({\it lower}; HXD/PIN). This clearly indicates different 
variability properties: count rates decreased 
above 3\,keV ({\it middle} and {\it bottom} panel), while it reached 
a delayed maximum $\sim$1.5 day from the start of the 
{\suzaku} observation for the 0.4$-$1\,keV band. 

\begin{figure}
\begin{center}
\includegraphics[angle=0,scale=.70]{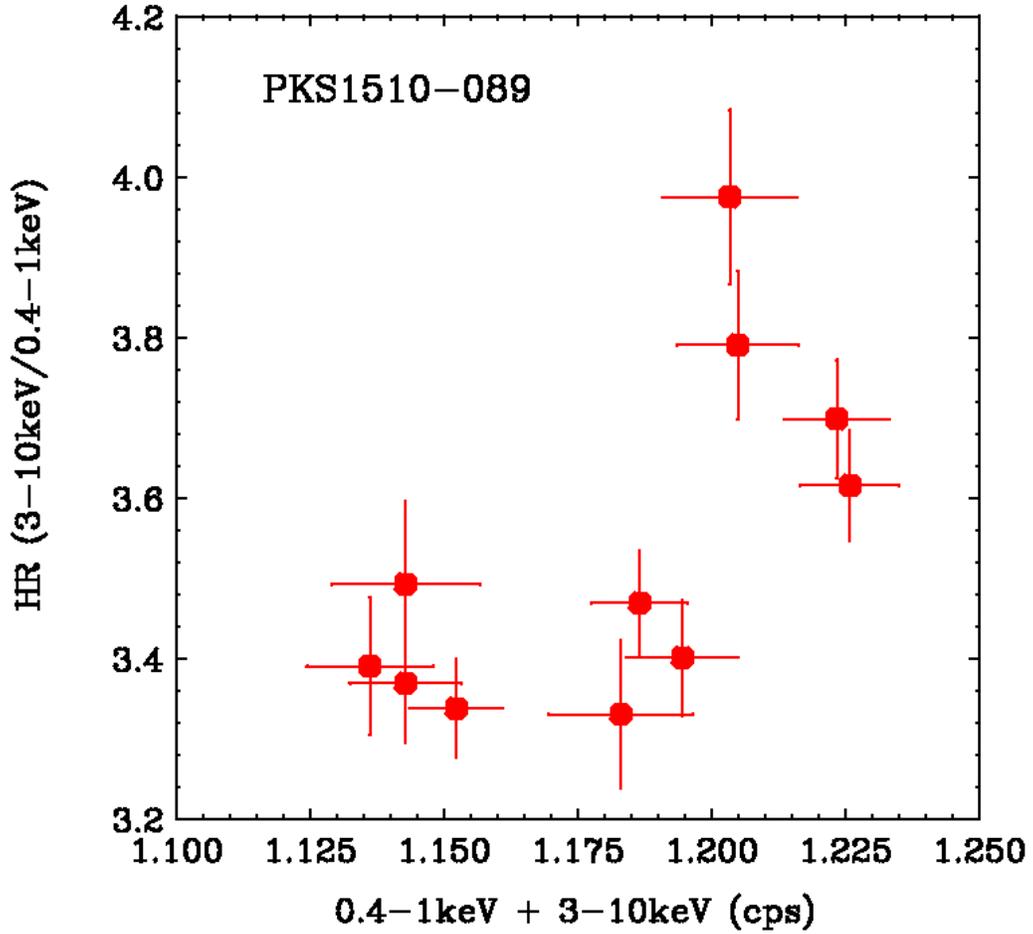}
\caption{Changes in the hardness ratio between 0.4$-$1\,keV and 3$-$10\,keV.
The hardness is defined as the 3$-$10\,keV count rate divided by the 
0.4$-$1\,keV count rate (XIS 0$-$3 summed).}\label{fig:suzakuhr}
\end{center}
\end{figure}

Spectral evolution during the observation is best illustrated as a 
correlation between the source brightness and the hardness ratio.
Figure~\ref{fig:suzakuhr} shows the count rate (sum of 0.4$-$1.0\,keV
and 3$-$10\,keV counts) 
versus hardness ratio, defined as the ratio of the XIS count rates 
at 3$-$10\,keV to those at the 0.4$-$1.0\,keV. This suggests a 
spectral evolution with the spectrum hardening as 
the source becomes brighter, although the correlation is rather loose, 
especially when the source is in the lower state of the activity 
(e.g., when the sum of 0.4$-$1\,keV and 3$-$10\,keV count rates is less than 
1.2\,ct\,s$^{-1}$). In addition, the hardness parameter is not a 
linear function of the source brightness, at least for the  
relatively short timescale of the {\suzaku} observation (i.e., within 
three days). Again, this may suggest that more than just one spectral 
component contributes to the observed X-ray emission of PKS~1510$-$089.

The XIS and HXD/PIN background subtracted spectra were fitted using 
XSPEC v11.3.2, including data within the energy range 0.3$-$50\,keV.
We binned the XIS spectra to a minimum of 400 counts per bin to 
enable the use of the $\chi^2$ minimization statistic. 
The Galactic absorption toward PKS~1510$-$089 is taken to be 
$N_{\rm H}$ = 7.88$\times$10$^{20}$\,cm$^{-2}$ (Lockman \& Savage 1995).
All errors are quoted at the 68.3\% (1$\sigma$) confidence level for 
the parameter of interest
unless otherwise stated. All the fits in this
section are restricted to the energy range 0.5$-$12\,keV (XIS 0, 2, 3: 
FI chips), 0.3$-$9\,keV (XIS 1: BI chip) and 12$-$50\,keV (HXD/PIN).
In the following analysis, we fixed the relative normalization of the 
XISs/PIN at 1.13, which is carefully determined from calibration 
using the Crab nebula, pointed at the XIS nominal position 
({\suzaku} internal report; JX-ISAS-SUZAKU-MEMO-2006-40 by M.~Ishida).

\begin{figure}
\begin{center}
\includegraphics[angle=0,scale=.70]{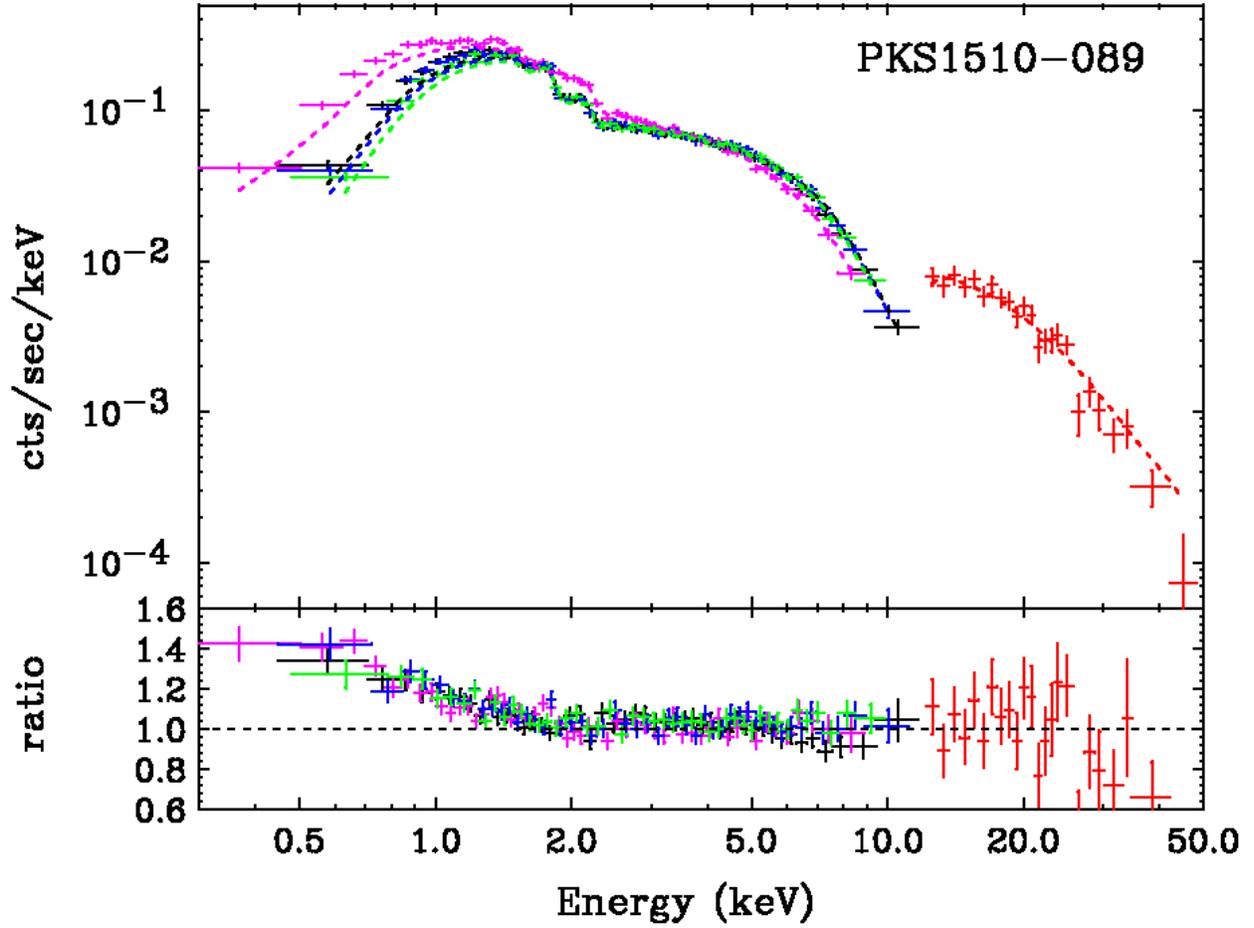}
\caption{$upper$ $panel$: The broadband (0.3$-$50\,keV; XIS0$-$3 + HXD/PIN) 
{\suzaku} spectrum of PKS~1510$-$089. The upper panel shows the data,
plotted against an absorbed power-law model of photon index $\Gamma$ =
1.2 and a column density 7.88$\times$10$^{20}$ cm$^{-2}$, fitted over the 
2$-$50\,keV band. The $lower$ $panel$ shows the data/model ratio residuals to 
this power-law fit. Deviations due to excess soft emission are clearly seen.}\label{fig:spfit}
\end{center}
\end{figure}

Figure~\ref{fig:spfit}  shows 4 XISs + HXD/PIN spectra with residuals to the 
best-fit power-law model determined using data between 
2 and 50\,keV. The residuals of Figure~\ref{fig:spfit} 
indicate that the spectrum 
exhibits significant soft excess emission below 2\,keV. If we 
model the overall X-ray spectrum between 0.3 and 50\,keV with a simple 
power-law function modified by the Galactic absorption, we obtain the 
best-fit photon index $\Gamma_{\rm high}$\,=\,1.30, but the fit is
statistically unacceptable with a 
$\chi^2$/d.o.f.\ of 705/585 
(Table~4). To represent the concave shape of the observed X-ray spectrum, 
we first consider a double power-law function (PL+PL) in which the soft 
X-ray excess is due to a steep power-law component with the 
photon index $\Gamma_{\rm low}$\,$\simeq$\,2.7. This provides
an acceptable fit with $\chi^2$/d.o.f.\,=\,536/585, although the
wavy structure still remains, especially below 1\,keV. 

\begin{figure}
\begin{center}
\includegraphics[angle=0,scale=.70]{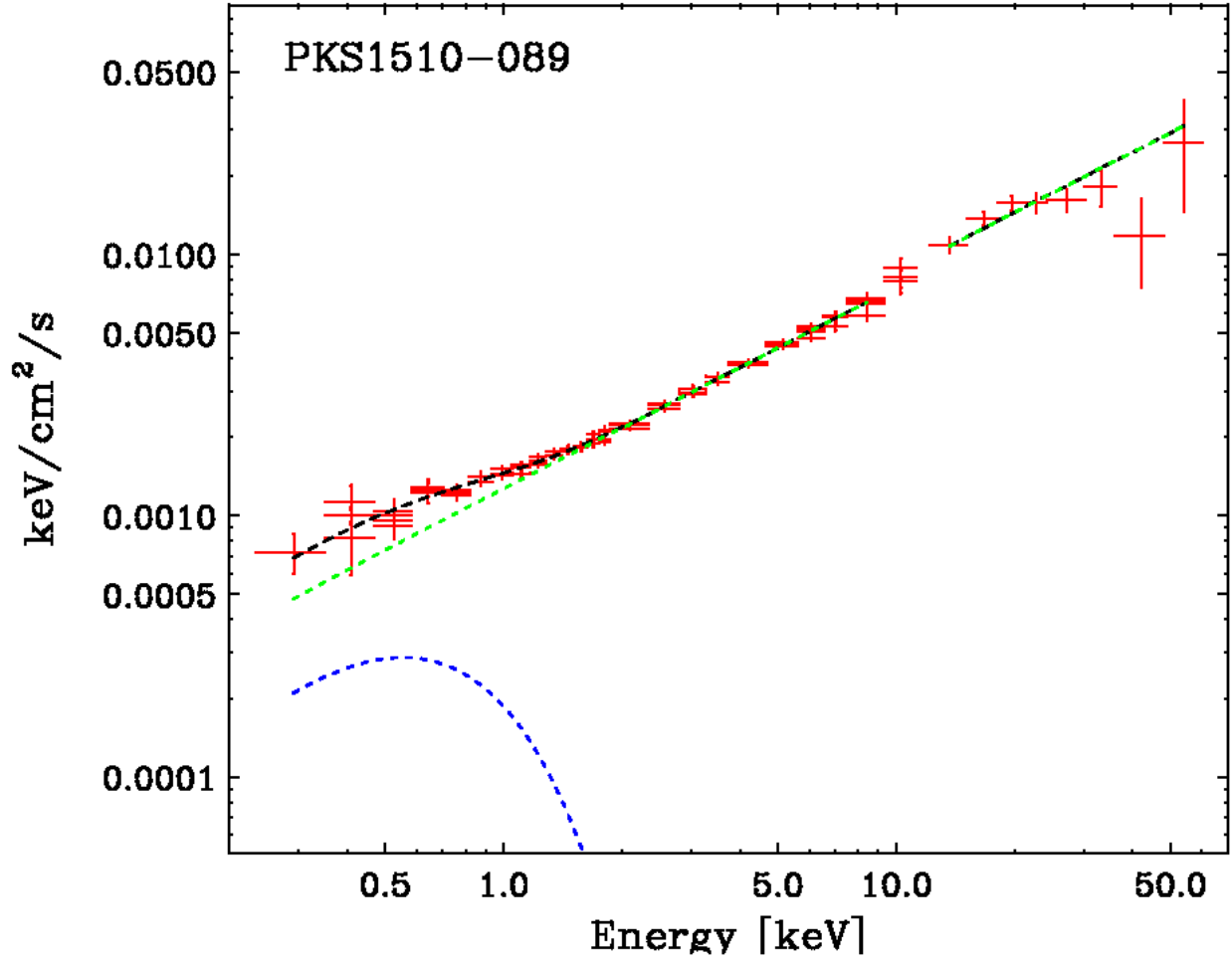}
\caption{The unfolded {\suzaku} spectrum between 0.3 and 50\,keV 
(in $\nu$$F_{\nu}$ space), plotted against the best-fit model composed 
of an absorbed power-law ($\Gamma$ = 1.2: green) plus 
disk black body emission (kT = 0.2\,keV: blue). The black line shows 
the sum of the model components.}\label{fig:nfn}
\end{center}
\end{figure}

Hence we considered an alternative fit consisting of 
a hard power-law function (PL) with $\Gamma_{\rm hard}$\,=\,1.23 and
a blackbody (BB) component or a disk blackbody (DB; Mitsuda et al.\ 1984) 
feature. Both models give 
similarly good representation of the data, with the best 
$\chi^2$/d.o.f.\,=\,515/585 for the latter (Table~4). The improvement 
of the $\chi^2$ statistic is significant at more than the 99.9 $\%$ 
confidence level when compared to the PL+PL model described above 
($\Delta$\,$\chi^2$\,$\simeq$\,20 for 583 d.o.f).
The temperature of the introduced thermal component is fitted as 
$kT$\,$\simeq$\,0.2\,keV. Figure~\ref{fig:nfn} shows an absorption corrected 
$\nu$$F_\nu$ spectrum deconvolved with this PL+DB model. 
The integrated luminosity of this blackbody-type 
emission is $L_{\rm BB}$\,$\simeq$\,(2.6 $\pm$ 0.2)$\times$10$^{44}$\,erg\,s$^{-1}$.

\subsection{\swift}

The deep {\suzaku} observations of PKS~1510$-$089 over three days 
(120\,ks in total) are well complemented by the {\swift}/XRT observation 
for monitoring the long-term variability of this source on the week-long 
scale. In addition, the first two observations 
made by {\swift} were simultaneous with {\suzaku} (see Table~1), 
and thus provide an important opportunity for the cross-calibration 
of results between the two instruments. Since the effective 
area of {\swift}/XRT is less than 10$\%$ of the XIS onboard 
{\suzaku} in the 0.5$-$10\,keV range, detailed spectral modeling is difficult 
using the {\swift} data alone. Furthermore, the average exposure 
for the {\swift} segment was only a few kilosecond, which was 
much less than the {\suzaku} exposure of 120\,ks. We therefore 
fit the XRT data simply with a power-law function in the energy 
band 0.3$-$10\,keV, modified by the Galactic absorption. 

\begin{figure}
\begin{center}
\includegraphics[angle=0,scale=.70]{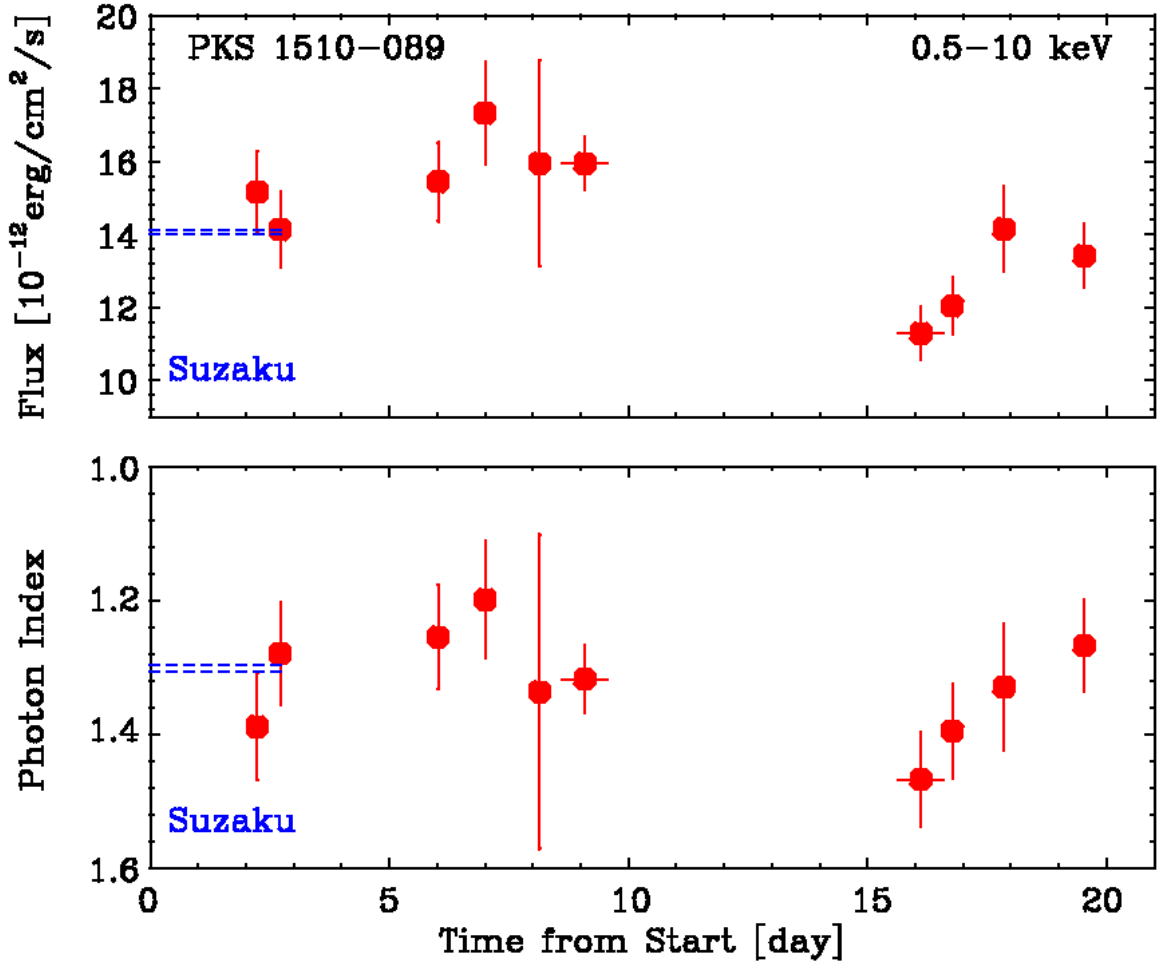}
\caption{Spectral variability of the {\swift}/XRT data during the 2006 August 
campaign. The observation time is measured from the start 
of the {\suzaku} observation, i.e., 2006 August 2, 09:31:29 UT, and the 
blue dashes show the best-fit parameters determined by {\suzaku}.
$upper$ $panel$: changes in the 0.5$-$10\,keV fluxes. Absorption corrected.
$lower$ $panel$: changes in the power-law photon index.}\label{fig:swlc}
\end{center}
\end{figure}

\begin{figure}
\begin{center}
\includegraphics[angle=0,scale=.70]{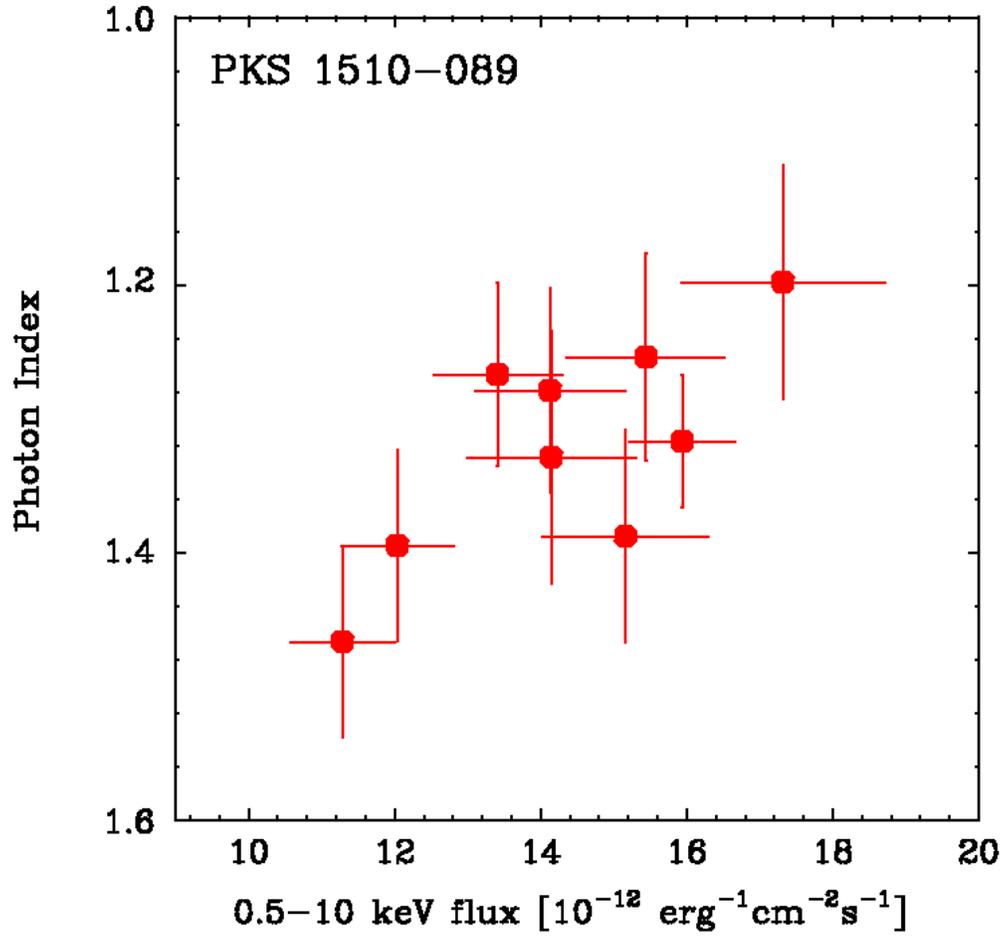}
\caption{Correlation of the 0.5$-$10\,keV flux versus photon index 
measured by the {\swift}/XRT. Data for observation \#5  
has been ignored in this plot simply due to the large uncertainties
(see error bars in Figure \ref{fig:swlc}).}\label{fig:swhr}
\end{center}
\end{figure}

Figure~\ref{fig:swlc} compares the variations of the X-ray flux and 
changes in the power-law spectral photon index as a function of time. 
Here the observation time is measured from the start of the 
{\suzaku} observation, i.e., 2006 August 2, 09:31 UT. Note the wide range of 
source variability (about a factor of two) on a week-scale, which was not 
observed with {\suzaku}. The blue lines in this figure show 
the time coverage and the best fit parameter determined by {\suzaku} 
(the two dashed lines show the 1\,$\sigma$  uncertainty of the {\suzaku} parameters). 
We confirm that the results obtained with {\suzaku} and {\swift} 
are perfectly consistent with each other. Figure~\ref{fig:swhr} shows the 
relation between the 0.5$-$10\,keV flux versus photon index 
measured by {\swift}/XRT (the data from August 10 were excluded as 
they have large statistical uncertainties due to the short exposure; 
see Table~1). Clearly the X-ray spectrum becomes harder when the 
source gets brighter. Such a trend is often observed in high-frequency 
peaked BL Lacs (e.g., Kataoka et al.\ 1999), but has not previously been 
observed so clearly in a quasar hosted blazar such as PKS~1510$-$089. 

\begin{figure}
\begin{center}
\includegraphics[angle=0,scale=.70]{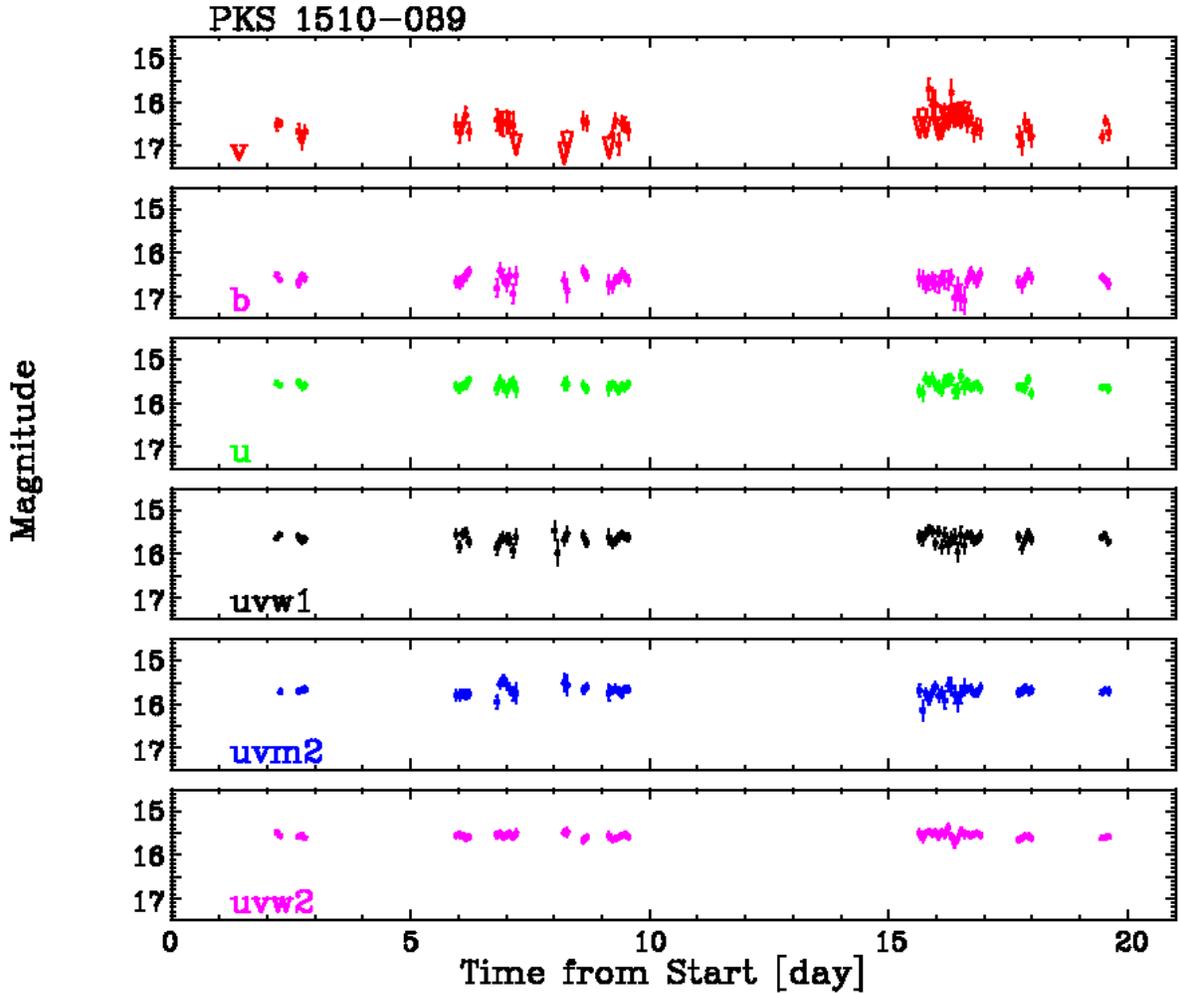}
\caption{The overall variability of the {\swift}/UVOT data during the 
2006 August campaign. From top to bottom; v-band, b-band, u-band, 
uvw1-band, uvm2-band, and uvw2-band. For more details about UVOT 
filters and their wavelength properties, see Poole et al., in preparation,   
and http://swift.gsfc.nasa.gov/docs/swift/about\_swift/uvot\_desc.html.}
\label{fig:uvlc}
\end{center}
\end{figure}

Thanks to the excellent sensitivity and wideband coverage of {\swift}/UVOT, 
even relatively short 1\,ks exposures provide the deepest UV-optical 
measurement ever reported for this source in the literature. 
To cover as much bandpass as possible, we used all the filters 
(v, b, u, uvw1, uvm2, uvw2) for all 10 observations (Table~1). 
The fluxes in each filter were corrected for galactic extinction 
following the procedure described in Cardelli, Clayton, \& Mathis
(1989): Galactic extinction with $R_V$\,=\,3.1 and $E_{B-V}$\,=\,0.097 
taken from Schlegel et al.\ (1998).  We note that the expected 
extinction value is somewhat higher than the one given by Burstein and Heiles 
(1982; $E_{B-V}$ between 0.06 and 0.09). We generated a list of 
the amount of extinction that needs to be accounted for in each filter,
$A_{\lambda}$\,=\,$E_{B-V}$$\times$[$a$ $R_{\rm V}$ + $b$] where 
$a$ and $b$ are constants summarized in Table~5. Resultant 
correction factors to each filters were 
$\times$1.32, 1.45, 1.56, 1.82, 1.91 and 2.09, respectively 
for v, b, u, uvw1, uvm2, uvw2 band filters. The extinction-corrected 
light curves thus produced are shown in Figure~\ref{fig:uvlc}. 
In contrast to the X-ray light curve (Figure~\ref{fig:swlc}), 
no significant variability was detected throughout the 
2006 August campaign.  

\begin{figure}
\begin{center}
\includegraphics[angle=0,scale=.70]{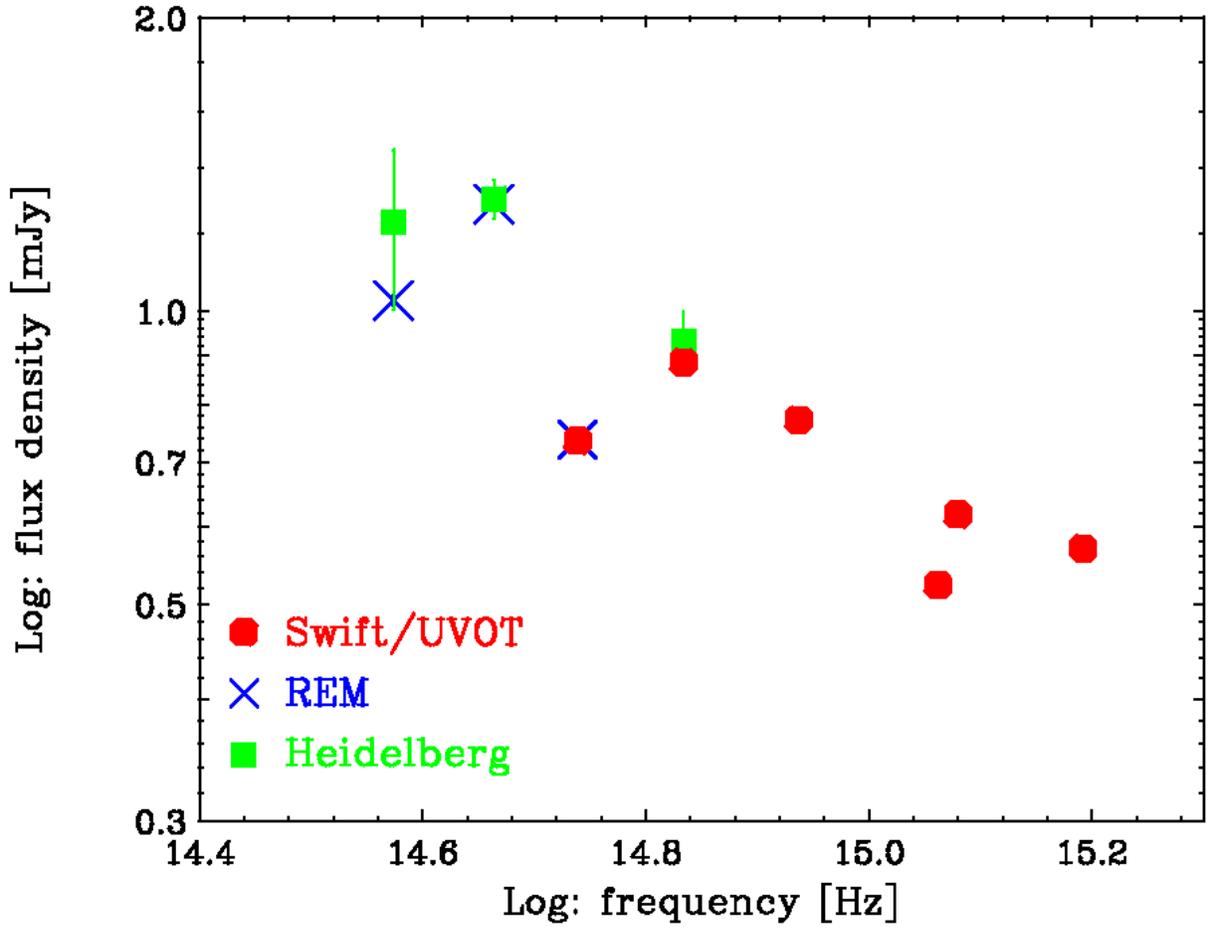}
\caption{Combined optical spectrum of PKS~1510$-$089 taken during the 
2006 August campaign. Data are corrected for the Galactic extinction 
using relations in Cardelli, Clayton \& Mathis (1989). 
Note the perfect consistency 
between {\swift}/UVOT, REM and Heidelberg telescopes.}\label{fig:opt}
\end{center}
\end{figure}

Finally, Figure~\ref{fig:opt} 
shows the combined optical/UV spectrum of PKS~1510$-$089 
taken during the campaign. {\it Red circles} show the average 
{\swift}/UVOT data, whereas {\it green} and {\it blue} points show data taken by 
REM and Landessternwarte, Heidelberg. Note the excellent agreement 
between {\swift}/UVOT and other telescopes at optical wavelengths.
The overall trend of the optical/UV continuum is that the 
flux density ($F_{\nu}$ in mJy) decreases with increasing frequency, 
such that $F_{\nu}$ $\propto$ $\nu^{-0.57}$. 
Also, it appears that the spectrum shows an interesting 
dip/discontinuity around $\nu$ $\sim$ $10^{14.7}$\,Hz and $10^{15.0}$\,Hz.
No discontinuity is evident in the combined infrared/optical/UV 
SED and line flux measurements (e.g., Mg~II) provided by 
Malkan \& Moore (1986).  However, strong line-emission
present in a single Swift/UVOT filter can produce an apparent
discontinuity.  The discrepancy may also be explained by the fact
that the observations reported by Malkan \&
Moore (1986) were obtained over several epochs.
 
\section{Discussion}

\subsection{Flat X-ray Continuum and Spectral Evolution}

In the previous sections we have presented temporal and spectral analyses 
of {\suzaku} and {\swift} observations of PKS~1510$-$089 in August 2006. 
The great advantage of using {\it all} 
the {\suzaku} and {\swift} instruments is that 
we can resolve the spectral evolution on different 
time scales, from hours ({\suzaku}) to weeks ({\swift}). 
In particular, our campaign provided the first detection of time 
variability as short as the day-scale in the hard X-ray 
energy band (12$-$40\,keV). During the {\suzaku} observations, 
PKS~1510$-$089 was in a 
relatively high state with an average flux of  
$F_{\rm 2-10 keV}$ $\sim$ 1.1$\times$10$^{-11}$\,erg\,cm$^{-2}$\,s$^{-1}$,
that gradually decreased by about 10$\%$ over the duration of the observation.
Historically, the flux observed with {\suzaku} is more than two times 
higher than that observed with {\it BeppoSAX} in 1998 
(5.2$\times$10$^{-12}$\,erg\,cm$^{-2}$\,s$^{-1}$) or with 
{\it ASCA} in 1996 (8.6$\times$10$^{-12}$\,erg\,cm$^{-2}$\,s$^{-1}$). 
{\swift}/XRT sampled a range of continuum fluxes 
during the 18 days of the campaign, detecting significant spectral evolution
with the photon index $\Gamma$ changing from 1.2 to 1.5. 

The observed photon index is significantly lower than that of 
radio-loud quasars ($<$$\Gamma$$>$ = 1.66 $\pm$ 0.07; Lawson et
al.\ 1992; Cappi et al.\ 1997) or radio-quiet quasars 
($<$$\Gamma$$>$ = 1.90 $\pm$ 0.11; Lawson et al.\ 1992; Williams et al.\ 
1992), and is more similar to the ones observed 
in high-redshift quasars. For example, 
in the sample of 16 radio-loud quasars at $z>2$ considered by 
Page et al.\ (2005), four sources have hard spectra with $\Gamma$\,=\,1.4.
(see also Tavecchio et al.\ 2000 for 0836+710 and Sambruna et al.\ 2006
for the {\swift} blazar J0746+2449 with $\Gamma$\,$\simeq$\,1.3). 
Such hard 
X-ray spectra pose a challenge to the `standard' shock models of particle
acceleration, because they imply a very flat electron 
energy distribution. As long as the X-ray emission is due to the low-energy
tail of the ERC spectral component, the photon index $\Gamma$\,=\,1.2 corresponds to
the electron energy distribution $N(\gamma)$\,$\propto$\,$\gamma^{-1.4}$, where
$\gamma$ is the Lorentz factor of the ultrarelativistic (radiating) electrons.

This may suggest that shocks --- if indeed they are 
responsible for accelerating the jet particles
--- can produce relativistic electrons with an energy spectrum much
harder than the `canonical' power-law distribution $N(\gamma)$ $\propto$
$\gamma^{-2}$, 
or that another mechanism energizes the electrons, at least at 
the low energies, $\gamma \leq 10$, typically involved in production of  the 
X-ray emission 
within the ERC model (e.g., Tavecchio et al.\ 2007). 
The latter possibility was discussed by 
Sikora et al.\ (2002) who assumed a double power-law form 
of the injected (`freshly
accelerated') electrons, with the break energy 
$\gamma_{\rm br}$ $\sim$ 10$^3$ corresponding 
to the anticipated threshold of diffusive 
shock acceleration\footnote{See in this context 
Stawarz et al.\ (2007) for the case of particle acceleration at
mildly-relativistic shocks in large-scale jets.}.
Below that energy, the electrons must be accelerated by a different 
mechanism, e.g., involving instabilities driven by 
shock-reflected ions (Hoshino et al.\ 1992) or magnetic reconnection
(Romanova \& Lovelace 1992). 
%
These `alternative' processes can possibly account for the electron 
distribution 
being harder than $\propto$ $\gamma^{-2}$.  
The X-ray emission of PKS~1510$-$089 discussed here 
provides direct constraints on this crucial low-energy population of 
ultrarelativistic electrons in quasar jets. At the other end of the 
spectrum of the Compton component, archival EGRET (and, 
in the future, {\it GLAST}) observations may be used to constrain the high-energy
tail of the accelerated electrons.

In this context, it is interesting to revisit the spectral evolution
detected with the {\swift} XRT in PKS~1510$-$089. In general, the trend 
established for FSRQs is that only little X-ray variability is observed 
on short time-scales of hours to days. Even on longer time-scales the 
X-ray variations in FSRQs are usually small, and the X-ray spectral shape 
is almost constant. The best example for such a behavior is 3C~279, 
where the X-ray slope changed only a little during the historical 
outburst in 1988, from $\Gamma$\,=\,1.70 $\pm$ 0.06 to 1.58 $\pm$ 0.03 
(Makino et al.\ 1989). However, exceptions were found recently in several 
distant quasars. For example, RBS 315 changed the X-ray spectral slope 
from $\Gamma$\,=\,1.3 to 1.5 between two observations separated by three 
years (Tavecchio et al.\ 2007). Also, the hard X-ray spectrum of 0836+710 
softened from $\Gamma$ $\simeq$ 1.4 (as measured by {\it BeppoSAX}; 
Tavecchio et al.\ 2000) to $\Gamma$ $\simeq$ 1.8 
({\swift} observations; Sambruna et al.\ 2007). 

Such a variability pattern may simply imply that the  
distribution of ultrarelativistic electrons at low energies changes 
(though not dramatically) in some sources.
Another explanation, however, is that this spectral shape remains 
roughly constant,
but that the amount of contamination from the soft excess emission varies, 
affecting the spectral fitting parameters at higher 
energies ($E$ $\ge$ 2\,keV).
In fact, we showed that the photon index became a little steeper 
($\Delta$$\Gamma$ $\simeq$ 0.1; Table~4) 
when we fit the spectrum with a simple power-law 
function. This suggests that spectral evolution of PKS~1510$-$089 may be 
explained by the soft excess emission being more important when the 
source gets fainter, and becoming almost completely `hidden' behind the 
hard X-ray power-law ($\Gamma_{\rm hard}$ $\simeq$ 1.2) when 
the source gets brighter.

\subsection{Modeling the overall SED}

\begin{figure}
\begin{center}
\includegraphics[angle=0,scale=.80]{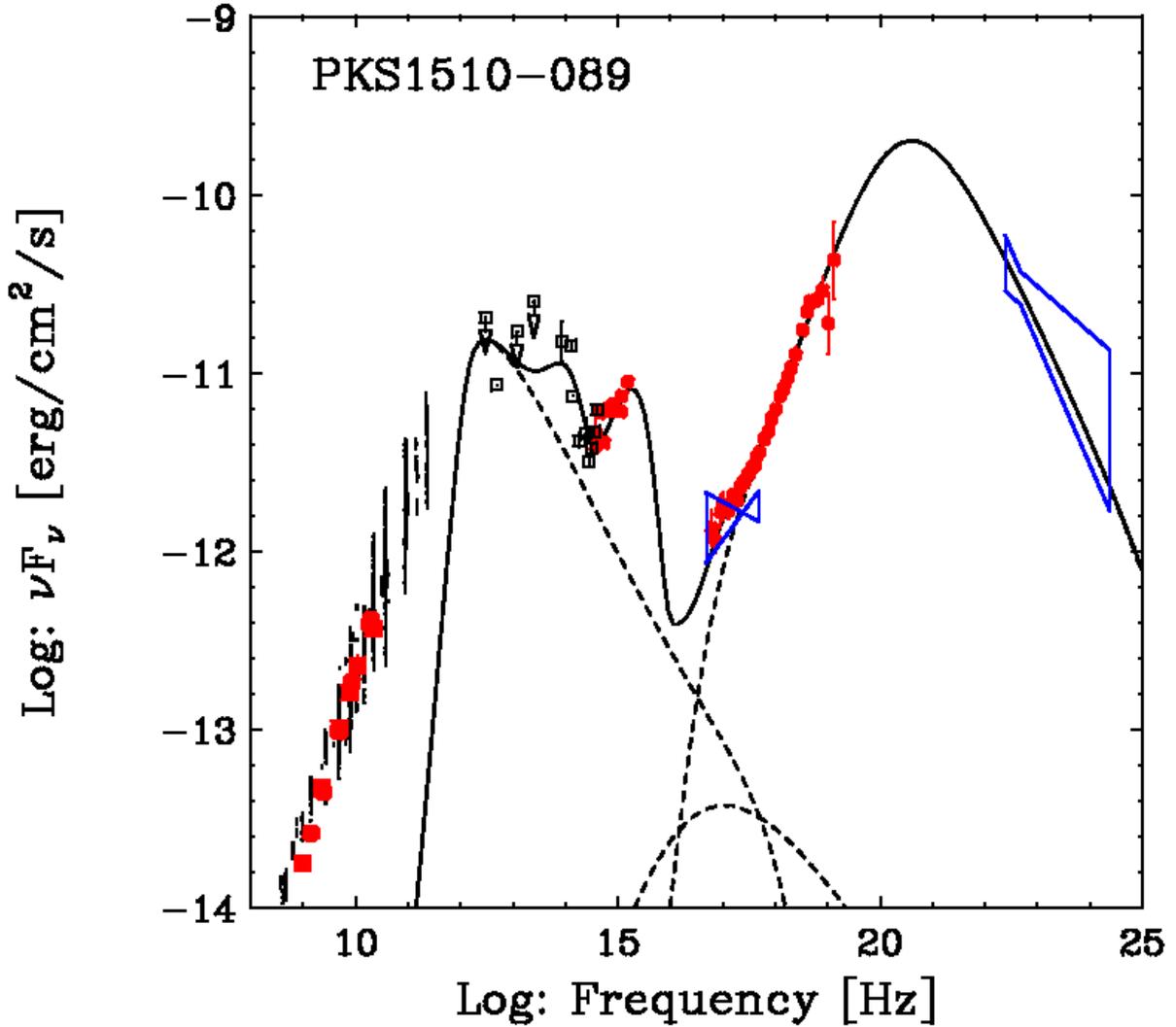}
\caption{Overall SED of PKS~1510$-$089 constructed with multiband data 
obtained during this campaign ($filled$ $circles$): radio (RATAN-600 and 
ATCA), optical ({\swift} UVOT, REM and Heidelberg), X-ray ({\suzaku}).  
Historical data taken from radio (NED and CATS), IR ({\it IRAS}; Tanner
 et al.\ 1996) optical (NED), soft X-ray ({\it ROSAT}; left bow-tie from 
Siebert et al.\ 1996) and $\gamma$-ray (EGRET; right bow-tie from 
Hartman et al.\ 1999) databases are also plotted as black points. 
The thick line shows the spectrum calculated 
using the jet emission model 
described in the  text, as a sum of various emission 
components (dotted lines; synchrotron, SSC, and ERC(IR); from left to 
right).  The input parameters for this model are: 
$\gamma_{\rm min}$ = 1, $\gamma_{\rm br}$ = 100, $\gamma_{\rm max}$
 = 10$^5$, $p$ = 1.35, $q$ = 3.25, $K_e$ = 0.9$\times$10$^{47}$ s$^{-1}$, 
$\Gamma_{\rm jet}$ = 20, $\theta_{\rm jet}$ = 0.05 rd, $\theta_{\rm obs}$ = 
0.05 rd, $r_{\rm sh}$ = 10$^{18}$ cm, $B = 1.3$ G, 
$r_{\rm ext}$ = 3.0$\times$$10^{18}$ cm, $L_{\rm ext}$ = 
3.7$\times$10$^{45}$ erg s$^{-1}$, and $h\nu_{\rm ext}$ 
= 0.2 eV. See Moderski, Sikora \& B\l a{\.z}ejowski (2003) for the 
definition of the input parameters.}\label{fig:MW1}
\end{center}
\end{figure}

Figure~\ref{fig:MW1} shows the overall spectral energy distribution (SED)
of PKS~1510$-$089 during the 2006 August campaign. Filled red circles
represent simultaneous data from the radio (RATAN-600 and ATCA),
optical ({\swift} UVOT, REM and Heidelberg),  and the X-ray ({\suzaku})
observations.  Historical data taken from the radio (NED
and CATS), FIR ({\it IRAS}, Tanner et al.\ 1996), optical (NED),
soft X-ray ({\it ROSAT}; Singh, Shrader \& George 1997) and $\gamma$-ray
(EGRET; Hartman et al.\ 1999) observations are also plotted as
black points or blue bow-ties. Figure~\ref{fig:MW1} implies that the
synchrotron component of PKS~1510$-$089 peaks most likely around
10$^{12-14}$\,Hz, while the excess at NIR frequencies
may be due to the starlight of the host, and the excess at FIR due to 
dust radiation from the nuclear torus. 
Meanwhile, our UVOT/REM/Heidelberg data show
the `rising' emission in the frequency range between 10$^{14.4}$ and
10$^{15.2}$Hz, with $\nu$$F_{\nu}$ $\propto$ $\nu^{0.43}$ ($\S$3.2.2).
As already suggested in
the literature (e.g., Malkan \& Moore 1986; Pian \& Treves 1993), 
this is most likely a manifestation of a strong 
`UV excess' (`blue-bump'), which is thought to be produced by
the accretion disk and/or corona near the central black hole of 
PKS~1510$-$089. Apparently, these optical/UV data do not join 
smoothly with the X-ray--to--$\gamma$-ray spectrum, which is due to
the non-thermal ERC jet radiation. Also note that the 
X-ray spectrum softens at low energies
due to the presence of soft excess emission, 
as suggested by the detailed spectral fitting in Figures~\ref{fig:spfit} 
and \ref{fig:nfn}.

To reproduce the overall SED of PKS~1510$-$089, 
we applied the numerical model {\it BLAZAR} developed in Moderski, Sikora \& 
B\l a{\.z}ejowski (2003), updated for the correct 
treatment of the Klein-Nishina
regime (Moderski et al.\ 2005). The code is based on a model in which 
the non-thermal flares in blazars are produced in thin shells 
propagating down a conical jet with relativistic velocities.
The production of non-thermal radiation is assumed to be dominated 
by electrons and positrons which are accelerated directly, rather than 
injected by pair cascades. The code traces the time evolution of 
the synchrotron and IC components, where both the synchrotron 
and external photons are considered as seed radiation fields
contributing to the IC process. We assumed that the electrons are injected 
by the shock formed at a distance $0.5r_{\rm sh}$, propagating with a Lorentz 
factor $\Gamma_{\rm sh}$, and decaying at $r=r_{\rm sh}$, 
and that the injection 
function takes  the broken power-law form  
\begin{equation} 
Q_{\gamma} = K_e \frac{1}{\gamma^p + \gamma_{\rm br}^{p-q}~\gamma^q} \, ,
\end{equation} 
where $K_e$ is the normalization factor, $p$ and $q$ are  
spectral indices of the injection function at the low and high energy 
limits, respectively, and $\gamma_{\rm br}$ is the break energy.

Co-moving energy density of the external radiation is approximated via 
\begin{equation} 
u'_{\rm ext} = \frac{4}{3} ~ \Gamma_{\rm sh}^2~ 
\frac{L_{\rm ext}}{4 \pi c r_{\rm ext}^2}~\frac{1}{1+(r/r_{\rm ext})^n} \, ,
\end{equation} 
where $L_{\rm ext}$ and $r_{\rm ext}$ are the total luminosity and 
the scale (spatial extent) of the considered external photon field, respectively, 
and $n$ $\geq$ 2. We investigated  radiation fields  
from both the dusty torus and the BLR and found 
that Comptonization of the former better reproduces the observed spectrum.
Our fit of PKS 1510-089 is shown on Figure~\ref{fig:MW1} and the model 
parameters are specified in the figure caption and in Table 6 (`Model A').
The presented model is a snapshot at the maximum of the flare 
which corresponds roughly to the distance $r_{\rm sh}$. 
Note that $r_{\rm sh} > r_{\rm BLR}$, but $r_{\rm sh}$ is still much below 
the radio photospheres
and the  observed flat-spectrum radio emission  originates from 
the superposition of  more distant, self-absorbed portions of the outflow.

\subsection{Energetics and Pair Content}

In order to derive the power and pair content of a jet, dynamics and structure
of the shock must be specified.
We adopt here internal shock scenario  and assume that 
shells with relativistic plasma represent regions enclosed
between the reverse and forward shock fronts. Such a structure is formed 
by colliding inhomogeneities propagating down the jet with different Lorentz 
factors. In this model, the light curves are produced by a sequence of shocks 
with a range of locations and lifetimes (Spada et al. 2000). Our fit
presented in Figure 10 shows the radiative output of the shock
operating over a distance range $\Delta r = 0.5 \times 10^{18}$cm, starting 
at $0.5 \times  10^{18}$cm and decaying at $r_{\rm sh}=10^{18}$cm. The amount
of electrons/positrons injected into shell by the end of the shock operation
is
\begin{equation} 
N_{e,{\rm inj}} = t_{\rm sh}' \int_{\gamma_{\rm min}} Q_{\gamma} d\gamma \simeq
{\Delta r \over c \Gamma_{\rm sh}} {K_e \over (p-1) \gamma_{\rm min}^{p-1}} \simeq
2.9 \times 10^{53} \, ,
\end{equation}
where $t_{\rm sh}' = \Delta r/(c\Gamma_{\rm sh})$ is the lifetime of the shock
as measured in the shock (discontinuity surface) rest frame.  

The electrons/positrons are accelerated/injected resulting in an average energy
$\bar \gamma_{\rm inj} = \int Q_{\gamma} \gamma d\gamma / \int Q_{\gamma} d\gamma
\simeq 22$ that follows from the model parameters as given in the caption of Figure 10
and in Table 6. Assuming that this energy is taken from protons, we have
the electron$+$positron to proton ratio 
\be {N_e \over N_p} = \eta_e {m_p (\bar\gamma_p -1) \over m_e
\bar\gamma_{\rm inj}} \, ,
\ee
where $\eta_e$ is the fraction of the proton thermal energy 
tapped by electrons and positrons.
The value of $\bar\gamma_p-1$, which actually represents efficiency of the 
energy dissipation, depends on properties and speeds of colliding 
inhomogeneities, and is largest if they have   same rest densities and  masses.
In this case, assuming $\Gamma_2 > \Gamma_1 \gg 1$,  
\be \bar\gamma_p-1 =
{(\sqrt{\Gamma_2/\Gamma_1}-1)^2 \over 2 \sqrt{\Gamma_2/\Gamma_1}} \, ,
\ee
and $\Gamma_{\rm sh}=\sqrt{\Gamma_1 \Gamma_2}$ (Moderski et al. 2004).

We assume  hereafter $\Gamma_1 = 10$ and $\Gamma_2 = 40$
and using Eqs. (4) and (5) obtain $N_e/N_p \sim 20 \eta_e$.
In this case the rest frame width of the shell by the end of the shock 
evolution is $\lambda' \simeq 0.4 (\Delta r / \Gamma_{\rm sh}) 
\simeq 1.1 \times 10^{16}$cm (see Appendix in Moderski et al. 2004), and 
density of electrons/positrons is  $n_e' = N_{e,{\rm inj}} /(\pi R^2 \lambda')$,
where $R=\theta_{\rm jet} r$ is the cross-sectional radius of a jet.
Using these  relations one can estimate energy flux carried by protons,
\be L_p \simeq n_p'  m_p c^3 \pi R^2 \Gamma_{\rm sh}^2 \sim
n_e' {N_p \over N_e} m_p c^3 \pi R^2 \Gamma_{\rm sh}^2 \sim 2.2 \times 10^{46}
\, (1-\eta_e)/\eta_e \, {\rm erg \, s}^{-1} \, .
\ee
This can be compared with the energy flux carried by magnetic fields
\be L_B \simeq {B^2 \over 8 \pi} c \pi R^2 \Gamma_{\rm sh}^2 \simeq 
6.3 \times 10^{45} \, {\rm erg \, s}^{-1} \, , \ee
and by electrons and positrons
\be L_e \simeq n_e' \bar\gamma_{\rm inj} m_e c^3 \pi R^2 \Gamma_{\rm sh}^2 \simeq
5.6 \times 10^{45} \, {\rm erg \, s}^{-1} \, , \ee
where $L_e$ is estimated  without taking into account radiative losses
of relativistic electrons/positrons  and therefore is overestimated 
by a factor $\sim 4$.

\subsection{Soft X-ray Excess}

\begin{figure}
\begin{center}
\includegraphics[angle=0,scale=.80]{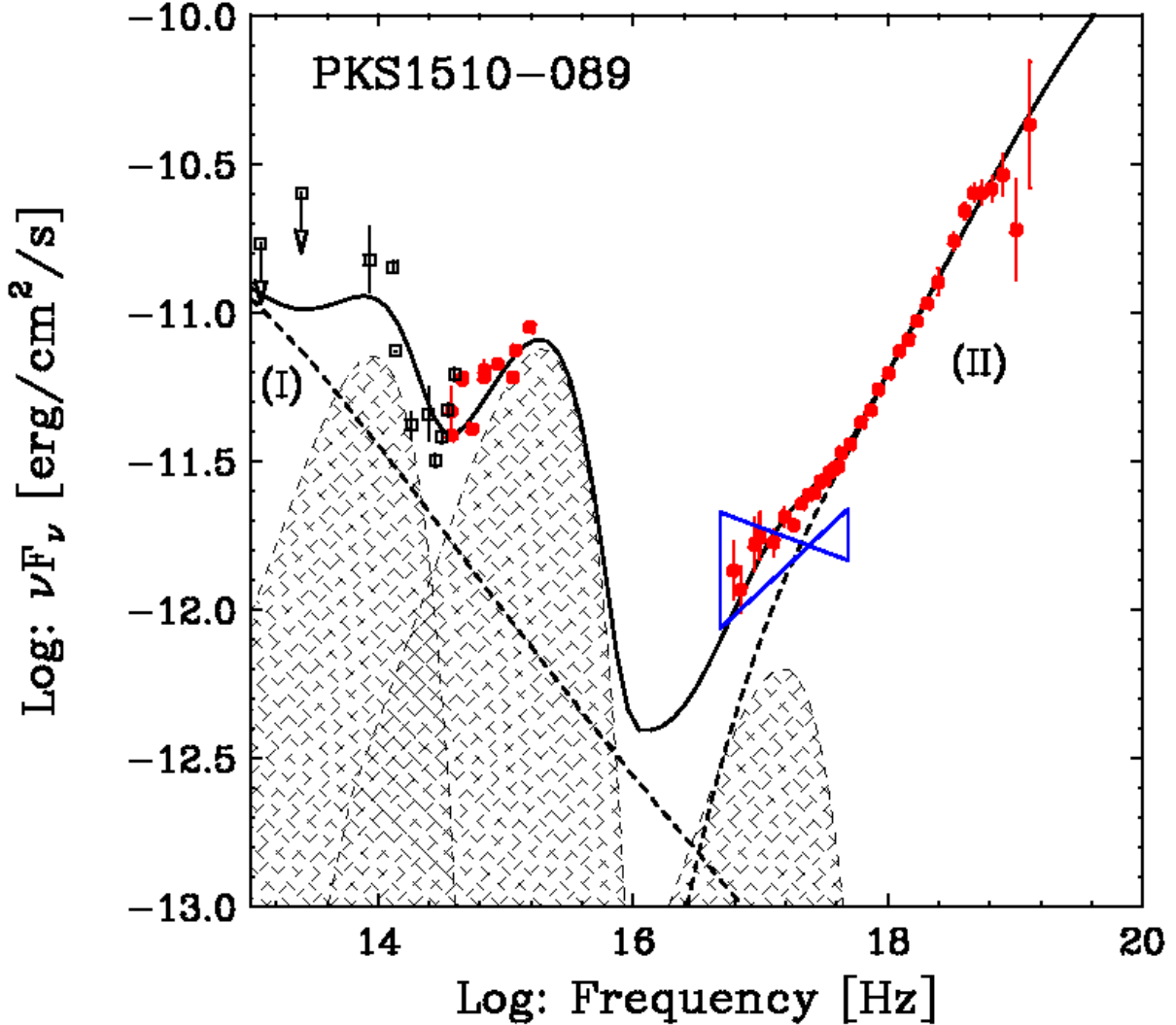}
\caption{A close-up of the PKS~1510$-$089 SED model presented in Figure 10
between the optical and the X-ray bands. Here are
added three blackbody-type humps: the left hump mimics an excess
 emission from the dusty torus as suggested by {\it IRAS} (Tanner et
 al. 1996), whereas the middle hump mimics the blue bump expected for this 
source from combined {\swift}/UVOT, REM, and Heidelberg data.
The right hump shows  the best fit blackbody-type emission of 
$kT$ $\simeq$ 0.2\,keV from the {\suzaku} fitting (Table~4). 
Dotted lines show (I) the Synchrotron and (II) the ERC components, 
respectively.A thick line shows sum of all the model components.}\label{fig:MW1_zoom}
\end{center}
\end{figure}

Figure~\ref{fig:MW1_zoom} shows in detail the optical--to--X-ray region of 
the SED. The hump on the left mimic an excess emission from the dusty 
torus as suggested by {\it IRAS} (Tanner et al.\ 1996) with a dust 
temperature of $kT$ $\simeq$ 0.2\,eV and $L_{\rm dust}$ $\simeq$
3.7$\times$10$^{45}$\,erg\,s$^{-1}$ (see also Table 6).
The hump on the middle is our attempt to account for the blue bump 
assuming an inner-disk temperature of $kT$ $\simeq$ 13\,eV and 
$L_{\rm disk}$ $\simeq$ 4$\times$10$^{45}$\,erg\,s$^{-1}$. 
We note here that the bolometric accretion luminosity is larger than this 
at least by a factor two, since more realistic models of accretion disks 
produce modified black body radiation, with extended high energy tails  
and additional contribution from more distant, cooler portions of a disk.
  
From the spectral fitting of the {\suzaku} data, we showed in $\S$3.1
and Table~4 that 
the soft X-ray excess may be represented either by a steep power-law 
($\Gamma_{\rm low}$ $\simeq$ 2.7) or a black-body--type emission of 
$kT$ $\simeq$ 0.2\,keV.  We investigate below whether such excess can be 
produced by a bulk Comptonization of external diffuse radiation by
cold inhomogeneities / density enhancements prior to their collisions. 
At $r > r_{\rm BLR}$ density of the diffuse external UV radiation
is very small, while bulk-Compton features from upscatterings of
dust infrared radiation falls into the invisible extreme-UV band.
However, if acceleration of a jet has already occurred at $r \le r_{\rm BLR}$,
upscattering of photons from broad-emission line region should lead to 
formation of bulk Compton features, with peaks located around
$\nu_{\rm BC,i} \sim {\cal D}_{\rm i} \Gamma_{\rm i} \nu_{\rm UV}/(1+z)$ and 
luminosities
\be L_{\rm BC,i} = {4 \over 3} c \sigma_{\rm T} u_{\rm BLR} \Gamma_{\rm i}^2 
{\cal D}_{\rm i}^4 N_{e,{\rm obs, i}} \, ,
\ee
where $i=1,2$, $u_{\rm BLR}$ is the energy density of the broad emission lines,
${\cal D}_{\rm i}$ is the Doppler factor, and  $N_{e,{\rm obs, i}}$ is the number 
of electrons and positrons contributing 
to the bulk-Compton radiation at a given instant (see Moderski et al. 2004).
Above formulas apply to cylindrical jets and 
must be modified if used as approximations for conical jets. For 
the conical jets the Doppler factor should  be replaced by the 'effective'
Doppler factor which for $\theta_{\rm obs} \le \theta_{\rm jet}$ is 
${\cal D}_{\rm i} = \kappa \Gamma_{\rm i}$, where $1< \kappa <2$.
For our model parameters
\be  N_{e,{\rm obs,1}} \simeq {N_{\rm inj} \over 2}  \,
{r_{\rm BLR} \over
\lambda_0 {\cal D}_{1}} \, , \ee
and
\be N_{e,{\rm obs,2}} \simeq {N_{\rm inj} \over 2}  \, 
{r_{\rm BLR} \over
\lambda_0 {\cal D}_{2}} \,
{\Gamma_{\rm sh}^2 \over 2\Gamma_2^2} 
\, , \ee
where $\lambda_0$ is the proper width (longitudinal size) of the cold 
inhomogeneities (see Appendix A3 in Moderski et al. 2004); the factor
$r_{\rm BLR} / (\lambda_0 {\cal D}_{\rm i})$ is the fraction of 
particles observed at a given instant and takes into account that 
the source is observed as being stretched to
the size $\lambda_0 {\cal D}_{\rm i}$ which is larger than $r_{\rm BLR}$; 
and the extra factor in the last equation, 
$ \Gamma_{\rm sh}^2 / (2 \Gamma_2^2)$, is the fraction
of particles enclosed within the Doppler beam.
With the above  approximations and $\kappa = 1.5$ our model predicts 
location of the bulk-Compton features at $\sim 1$ keV and $\sim 18$ keV,
and luminosities of $\sim 2 \times 10^{44}\, {\rm erg \, s}^{-1}$
and $2 \times 10^{46}\,{\rm erg \, s}^{-1}$, respectively.
Comparing these luminosities with luminosities of nonthermal radiation 
we  conclude that within the uncerainties regarding the details 
of the jet geometry and model parameters,
bulk-Compton radiation produced by slower 
inhomogeneities is sufficiently luminous to be responsible 
for the soft X-ray excess observed by {\suzaku}, while the faster one 
can be tentatively identified with a small excess at $ \sim 18$ keV 
seen in Fig. 11.

\subsection{On Alternative Models}

\noindent

Very hard X-ray spectrum measured by {\suzaku} --- with $\Gamma_{\rm X-ray} < 1.5$ --- 
excludes models in which 
X-rays are produced by synchrotron radiation of the secondary 
ultrarelativistic population of electrons/positrons predicted by hadronic 
models, whereas the  
large luminosity ratio of the high to low energy components
challenges the SSC models via enforcing magnetic 
fields much below the equipartition value (Moderski \& Sikora 2007).
No such constraints apply to ERC models, provided jets are sufficiently 
relativistic. Furthermore, in the comoving frame of a jet moving with 
a Lorentz factor $\ge 6$, the 
 energy density of seed photon fields is dominated by
broad emission lines or infrared radiation of dust (depending on
a distance of the source from the black hole), rather than by
locally produced synchrotron radiation (Sikora et al. 1994).
On the other hand, leptonic content of jets in the SSC models can be much 
smaller than in the ERC models, and therefore such models can easily avoid 
production of bulk-Compton features which in the context of ERC models are 
predicted to be prominent (Moderski et al. 2004) but so far were not
observationally confirmed. However, that concern applies only to 
ERC models with the seed photons from the BLR. Production
of similar high energy spectra but at larger distances where 
external diffuse radiation field is 
dominated by near/mid infrared radiation involves smaller leptonic
content and therefore weaker bulk-Compton features (Sikora et al.,
in preparation).  Furthermore, the predicted bulk-Compton features can be 
weaker in a scenario where the jet may still be in 
the acceleration phase while traversing the 
BLR (see Komissarov et al. 2007 and refs. therein).

If the latter is the case, the soft X-ray excess in
PKS~1510$-$089 cannot originate from the bulk-Compton process. 
A possible alternative origin of the observed soft X-ray excess may be
then provided by central regions of the accertion flow (Done \& Nayakshin 2007), as
seems to be the case in many non-blazar AGNs (Crummy et al. 2007 and refs. therein; 
see also Figure 6 in Laor et al. 1997).

Yet another possibility for the origin of the observed soft X-ray excess 
would be due to a more significant contribution of the 
SSC component in a frequency range between the synchrotron and ERC peaks, that
would remain essentially invisible in the UV and hard X-ray bandpasses. In fact,
we find that the collected data for PKS~1510$-$089 are consistent with 
another set of model parameters than discussed previously (\S~4.2-4.3), for which the soft X-ray
excess is due to a combination of (I) the tail of the synchrotron component, 
(II) the ERC, and (III) the SSC emission falling in the soft X-ray band
(Figures 12 and 13; see also parameters of `Model B' in Table 6). Some discrepancies seen at
10$^{17}$$-$10$^{18}$\,Hz are probably due to the mismatch of spectral 
slopes between the expected index of the SSC component 
($\Gamma$ $\simeq$ 2) and the observed, steep   
excess emission ($kT$ $\simeq$ 0.2\,keV or $\Gamma_{\rm low}$ $\simeq$ 2.7). 
A steeper power-law contribution might naturally be provided 
just by  the high-energy tail of the synchrotron emission, as often observed 
in low-frequency peaked blazars 
(e.g., Madejski et al.\ 1999; Tanihata et al.\ 2000; Tagliaferri et al.\ 2000).
However, we note that extrapolation of this power-law 
to lower frequencies overpredicts 
the observed IR/optical continuum about an order of magnitude.  Still, even
if any of the above models proves to be more appropriate as a description 
of the soft excess, the results obtained in Sec. 4.4 for the 
bulk-Compton features should be considered as upper limits which would then 
imply that the Lorentz factor of the jet is lower in the BLR region 
than at distances where the nonthermal radiation is produced.

\section{Conclusions}

We have presented a detailed analysis of the data for the powerful 
$\gamma$-ray -- emitting quasar PKS~1510$-$089 obtained with 
{\suzaku}, {\swift} XRT/UVOT, and ground-based optical (REM, Heidelberg)  
and radio (RATAN-600, ATCA) telescopes. Observations were conducted in 
2006 August as an intensive multiwavelength three-week--long campaign. 
An excellent broadband spectrum of the source was uniquely constructed,
covering ten decades in frequency, from 10$^9$\,Hz to 10$^{19}$\,Hz. 
Our major findings are as follows:

\renewcommand\theenumi{(\roman{enumi})}
\begin{enumerate}
\item Deep {\suzaku} observations indicate moderate 
X-ray variability of PKS~1510$-$089
on the time-scale of days, although differing in nature between 
the low (0.4$-$1\,keV) and high ($\ge$\,3\,keV) energy bands.
\item The X-ray spectrum of PKS~1510$-$089 is well represented 
by an extremely hard power-law (photon index $\Gamma$ $\simeq$ 1.2) 
augmented by a blackbody-type component (temperature $kT$ $\simeq$ 0.2\,keV) 
that accounts for the excess emission below 1\,keV.
\item {\swift}/XRT observations reveal significant 
spectral evolution of the X-ray emission on the timescale of
a week: the X-ray spectrum becomes harder as the source gets brighter.
\item  Using ERC model we found that best fit of the broad-band
spectrum is obtained by assuming that nonthermal radiation is produced 
at $r>r_{\rm BLR}$ where external diffuse radiation field is dominated by 
IR radiation of hot dust.
\item The model predicts that the electron to proton ratio 
$N_e/N_p \sim 10$ and that the power 
of the jet is dominated by protons.
\item Prior to collisions and formation of shocks, density inhomogeneities 
interact with the BLR light and produce bulk-Compton features which
tentatively can
be identified with the features observed in the {\suzaku} spectra: 
soft X-ray excess seen below $\sim 1$ keV, 
and another, marginally significant spectral feature at $\sim 18$ keV. 
\item Alternatively, the soft X-ray excess 
can be explained as a contribution of SSC component or it  can just be 
identified with the 
soft X-ray excess often observed in the non-blazar AGN.
\end{enumerate}

\acknowledgments

We are grateful to all the {\suzaku} members who helped us in analyzing
the data. We also thank all the {\swift} members (especially Neil Gehrels 
who approved the target) for performing the ToO observations.
We thank an anonymous referee for his/her helpful comments which
helped clarify many of the issues presented in this paper.
This work is sponsored  at PSU by NASA contract NAS 5-00136.
G.T.\ and D.I.\ thank the REM team  for the support 
received during the observations. \mbox{RATAN--600} observations 
were partly supported by the Russian Foundation for Basic Research 
(projects 05--02--17377). 
The ATCA is part of the Australia Telescope
which is funded by the Commonwealth of Australia for operation as a
National Facility managed by CSIRO.
P.G.E.\ thanks Mark Wieringa for assistance with the ATCA observations.
The authors made use of the database 
CATS (Verkhodanov et al.\ 1997) of the Special Astrophysical Observatory. 
This research has made use of the NASA/IPAC Extragalactic Database (NED) 
which is operated by the Jet Propulsion Laboratory, California 
Institute of Technology, under contract with the National Aeronautics 
and Space Administration.  
J.K.\ and N.K.\ acknowledge support by JSPS KAKENHI 
(19204017/14GS0211). G.M. and \L .S. acknowledge the support by the 
Department of Energy contract to SLAC no.\ DE-AC3-76SF00515.  
M.S.\ and R.M.\  were partially supported by 
MEiN grant 1-P03D-00928. G.M. acknowledges 
the support via the NASA Suzaku grant no. NNX07AB05G.   
Y.Y.K is a Research Fellow of the Alexander von Humboldt Foundation. 
\L .S. was supported by MEiN through the research project 1-P03D-003-29 
from 2005 to 2008.

\begin{table}
  \caption{2006 {\suzaku}/{\swift} observation log of PKS~1510$-$089.}\label{tab:first}
  \begin{center}
    \begin{tabular}{llll}  
    \tableline
    Instr & &  & \\
    start (UT) & stop (UT) & Exposure (ks)& Exposure (ks)\\
    \tableline\tableline
    {\suzaku} &  & XIS & HXD\\
    Aug 02  09:31 & Aug 05 06:06 & 119.2 & 93.4\\  
    \tableline
    {\swift} & & XRT  & UVOT (v/b/u/uvw1/uvm2/uvw2)\\
    Aug 04  14:09 & Aug 04 16:10 & 2.6 & 0.10/0.13/0.13/0.27/0.11/0.55\\  
    Aug 05  01:10 & Aug 05 04:35 & 2.1 & 0.09/0.09/0.09/0.17/0.23/0.35\\ 
    Aug 08  05:07 & Aug 08 14:54 & 2.2 & 0.07/0.07/0.07/0.14/0.16/0.27\\  
    Aug 09  04:49 & Aug 09 14:31 & 1.6 & 0.07/0.07/0.07/0.14/0.14/0.28\\  
    Aug 10  09:47 & Aug 10 16:13 & 0.5 & 0.05/0.05/0.05/0.12/0.11/0.21\\
    Aug 11  00:15 & Aug 11 22:51 & 4.3 & 0.14/0.14/0.14/0.29/0.37/0.58\\     
    Aug 18  00:57 & Aug 18 23:30 & 3.2 & 0.13/0.13/0.13/0.27/0.25/0.54\\     
    Aug 19  01:02 & Aug 19 07:37 & 2.6 & 0.11/0.11/0.11/0.22/0.28/0.46\\  
    Aug 20  02:46 & Aug 20 09:20 & 2.2 & 0.09/0.09/0.09/0.18/0.21/0.37\\  
    Aug 21  20:32 & Aug 22 23:55 & 3.0 & 0.13/0.13/0.13/0.25/0.35/0.52\\  
    \tableline
    \end{tabular}
  \end{center}
\end{table}

\begin{table}
  \caption{Optical observation log of PKS~1510$-$089.}\label{tab:first}
  \begin{center}
    \begin{tabular}{rccccc}  
    \tableline
    Instr & Filter & start (UT) & stop (UT) & exp. [s] &  Magnitude$^a$\\
    \tableline\tableline
    REM & V & 2006 Aug 20 23:14:21 & 2006 Aug 21 00:00:52 & 3000 & 16.88
     $\pm$0.02$^b$ \\  
    REM & R & 2006 Aug 19 23:13:21 & 2006 Aug 19 23:59:53 & 3000 & 16.10 $\pm$0.01 \\  
    REM & I & 2006 Aug 21 00:06:22 & 2006 Aug 21 00:52:54 & 3000 & 16.10 $\pm$0.01 \\  
    \tableline
    Heidelberg & B & 2006 Aug 01 20:25:12 & 2006 Aug 01 21:19:51 & 480 & 16.95
$\pm$0.08 \\  
    Heidelberg & R & 2006 Aug 01 20:30:22 & 2006 Aug 01 21:24:23 & 480 & 16.15
$\pm$0.05 \\  
    Heidelberg & I & 2006 Aug 01 20:35:09 & 2006 Aug 01 21:29:14 & 360 & 15.90
$\pm$0.20 \\  
    \tableline
    \end{tabular}
   \tablenotetext{a}{Observed magnitude for each observation using specific
   filters (Galactic extinction not corrected).
$^b$: On August 19th PKS~1510$-$089 showed a very fast rise ($\Delta m_{\rm R}$ 
$\simeq$ 0.6) in less than one hour.}
  \end{center}
\end{table}

\begin{table}
  \caption{Radio observation log of PKS~1510$-$089.}\label{tab:first}
  \begin{center}
    \begin{tabular}{rccc}  
    \tableline
    Instr & Frequency [GHz] & Observation Time & Flux density [Jy]$^a$\\
    \tableline\tableline
    RATAN & 1.0 & 2006 Aug 9 $-$ Aug 23 & 1.88$\pm$0.07$^b$\\ 
    RATAN & 2.3 & 2006 Aug 9 $-$ Aug 23 & 2.01$\pm$0.05$^b$\\  
    RATAN & 4.8 & 2006 Aug 9 $-$ Aug 23 & 2.08$\pm$0.03$^b$\\  
    RATAN & 7.7 & 2006 Aug 9 $-$ Aug 23 & 2.09$\pm$0.04$^b$\\  
    RATAN & 11.1 & 2006 Aug 9 $-$ Aug 23 & 2.05$\pm$0.21$^b$\\
    RATAN & 21.7 & 2006 Aug 9 $-$ Aug 23 & 1.68$\pm$0.08$^b$\\  
   \tableline
    ATCA & 1.4 & 2006 Jul 11  & 1.93$\pm$0.11\\ 
    ATCA & 2.3 & 2006 Jul 11  & 1.83$\pm$0.10\\ 
    ATCA & 4.8 & 2006 Jul 11  & 1.85$\pm$0.09\\ 
    ATCA & 8.6 & 2006 Jul 11  & 1.97$\pm$0.10\\ 
    ATCA & 1.4 & 2006 Aug  4  & 1.85$\pm$0.14\\ 
    ATCA & 2.4 & 2006 Aug  4  & 1.84$\pm$0.10\\ 
    ATCA & 4.8 & 2006 Aug  4  & 2.02$\pm$0.10\\ 
    ATCA & 8.6 & 2006 Aug  4  & 2.12$\pm$0.11\\ 
    ATCA & 18.5 & 2006 Aug 4  & 2.08$\pm$0.11\\ 
    ATCA & 19.5 & 2006 Aug 4  & 2.11$\pm$0.12\\ 
   \tableline
    \end{tabular}
   \tablenotetext{a}{The flux density errors presented do not include the error of the absolute radio flux density scale. See its estimate in Baars et al.\ 
(1977) and Ott et al.\ (1994).}  
   \tablenotetext{b}{RATAN flux densities averaged over the period
   August 9--23, 2006. }  
  \end{center}
\end{table}

\begin{table}
\small{
  \caption{Results of the spectral fits to the 0.3$-$50\,keV {\suzaku}
 spectrum with different models.}\label{tab:first}
  \begin{center}
    \begin{tabular}{rcccccccc}  
    \tableline
    Model$^a$ & $N_{\rm H}$$^b$ & $\Gamma_{\rm high}$$^c$ & $\Gamma_{\rm low}$$^d$ &
     $kT$$^e$ & $F_{\rm 0.5-10\,keV}$$^f$ & $F_{\rm 2-10\,keV}$$^f$ & $F_{\rm 10-50\,keV}$$^f$ & $\chi^2$/d.o.f \\
    \tableline\tableline
    PL & 7.88 & 1.30$\pm$0.01 & ... & ... & 14.1$\pm$0.1&
     10.8$\pm$0.1 & 33.7$\pm$0.4 & 705/585 \\ 
    PL+PL & 7.88 & 1.19$\pm$0.02 & 2.76$\pm$0.15 & ... & 14.4$\pm$0.1&
     11.1$\pm$0.1 & 39.7$\pm$0.5 & 536/583 \\ 
    \tableline
    PL+BB & 7.88 & 1.24$\pm$0.01 & ... & 0.16$\pm$0.02 & 14.2$\pm$0.1&
     11.1$\pm$0.1 & 38.3$\pm$0.4 & 519/583 \\ 
    PL+DB & 7.88 & 1.23$\pm$0.01 & ... & 0.23$\pm$0.02 & 14.4$\pm$0.1&
     11.1$\pm$0.1 & 38.2$\pm$0.4 & 515/583 \\ 
   \tableline
    \end{tabular}
   \tablenotetext{a}{Spectral fitting models. PL: power-law function, 
PL+PL: double power-law functions, PL+BB: power-law + blackbody model, 
PL+DB: power-law + disk blackbody model.}
   \tablenotetext{b}{Galactic absorption column density in units of 10$^{20}$ 
cm$^{-2}$.}
   \tablenotetext{c}{Differential spectral photon index.}
   \tablenotetext{d}{Differential spectral photon index at low energy X-ray band, when fitted with a double power-law functions.}
   \tablenotetext{e}{Temperature at inner disk radius in keV, fitted with disk blackbody model by Mitsuda et al. (1984).}
   \tablenotetext{f}{Flux in units of 10$^{-12}$ erg cm$^{-2}$ s$^{-1}$.}
  \end{center}
}
\end{table}

\begin{table}
  \caption{Correction factors for the Galactic extinction in UV and optical 
filters.}\label{tab:first}
  \begin{center}
    \begin{tabular}{rccccc}  
    \tableline
    Filter & $\lambda$ [nm]$^a$ & $a^{\dagger}$ & $b^{\dagger}$ & $A_{\lambda}^{\dagger}$ & $C_{\rm cor}$$^b$  \\
    \tableline\tableline
    I & 800 & 0.7816 & $-$0.5707  & 0.18 & 1.18\\ 
    R & 650 &0.9148 & $-$0.2707  & 0.25 & 1.26\\ 
    \tableline   
    v & 547  &1.0015 & 0.0126 & 0.30 & 1.32\\ 
    b & 439 &0.9994 & 1.0171 & 0.40 & 1.45\\ 
    u & 346 &0.9226 & 2.1019 & 0.48 & 1.56\\
    uvw1 & 260 & 0.4346 & 5.3286 & 0.65 & 1.82\\ 
    uvm2 & 249 & 0.3494 & 6.1427 & 0.70 & 1.91\\ 
    uvw2 & 193 & $-$0.0581 & 8.4402 & 0.80 & 2.09\\
    \tableline  
   \end{tabular}
   \tablenotetext{a}{Center wavelength for each optical/UV filters.}
   \tablenotetext{\dagger}{Parameters for calculating Galactic
   extinction for optical and UV filters, calculated according to the 
   prescription in Cardelli, Clayton \& Mathis (1989). The Galactic reddening 
   was taken from Schlegel, Finkbeiner, \& Davis (1998).}
   \tablenotetext{b}{Correction factor for Galactic extinction.}
  \end{center}
\end{table}

\newpage

\begin{table}
  \caption{The input parameters for modelling of the non-thermal emission of PKS~1510$-$089.}\label{tab:modelfit}
  \begin{center}
    \begin{tabular}{lcc}  
    \tableline\tableline
    Parameter & Model A & Model B \\
    \tableline\tableline
    minimum electron Lorentz factor $\gamma_{\rm min}$ & $1$ & $1$ \\
    break electron Lorentz factor $\gamma_{\rm br}$ & $100$ & $150$ \\
    maximum electron Lorentz factor $\gamma_{\rm max}$ & $10^5$ & $10^5$ \\
    low-energy electron spectral index $p$ & $1.35$ & $1.35$ \\
    high-energy electron spectral index $q$ & $3.25$ & $3.25$ \\
    normalization of the injection function $K_{e}$ & $0.9 \times 10^{47}\,{\rm s^{-1}}$ & $1.7 \times 10^{47}\,{\rm s^{-1}}$ \\
    bulk Lorentz factor of the emitting plasma $\Gamma_{\rm jet}$ & $20$ & $20$ \\
    jet opening angle $\theta_{\rm jet}$ & $0.05$\,rad & $0.02$\,rad \\
    jet viewing angle $\theta_{\rm obs}$ & $0.05$\,rad & $0.05$\,rad \\
    scale of the emission zone $r_{\rm sh}$ & $10^{18}$\,cm & $10^{18}$\,cm \\
    jet magnetic field intensity $B$ & $1.3$\,G & $0.86$\,G \\
    scale of the dominant external photon field $r_{\rm ext}$ & $3.0 \times 10^{18}$\,cm & $3.0 \times 10^{18}$\,cm \\
    luminosity of the external photon field $L_{\rm ext}$ & $3.7 \times 10^{45}\,{\rm erg\,s^{-1}}$ & $3.7 \times 10^{45}\,{\rm erg\,s^{-1}}$ \\
    photon energy of the external photon field $h \nu_{\rm ext}$ & $0.2$\,eV & $0.2$\,eV \\
    \tableline
    total energy of radiating electrons $E_{e}$ & $1.3 \times 10^{48}$\,erg & $3.1 \times 10^{48}$\,erg \\
    comoving electron energy density $u'_{e}$ & $0.015\,{\rm erg \, cm^{-3}}$ & $0.022\,{\rm erg \, cm^{-3}}$ \\
    equipartition magnetic field $B_{\rm eq}$ & $0.6$\,G & $2.4$\,G \\
    kinetic luminosity of radiating electrons $L_{e}$ & $1.4 \times 10^{45}\,{\rm erg \, s^{-1}}$ & $3.3 \times 10^{45}\,{\rm erg \, s^{-1}}$ \\
    \tableline\tableline   
    soft X-ray excess & bulk-Compton & SSC \\
    \tableline\tableline
   \end{tabular}
  \end{center}
\end{table}

\begin{figure}
\begin{center}
\includegraphics[angle=0,scale=.80]{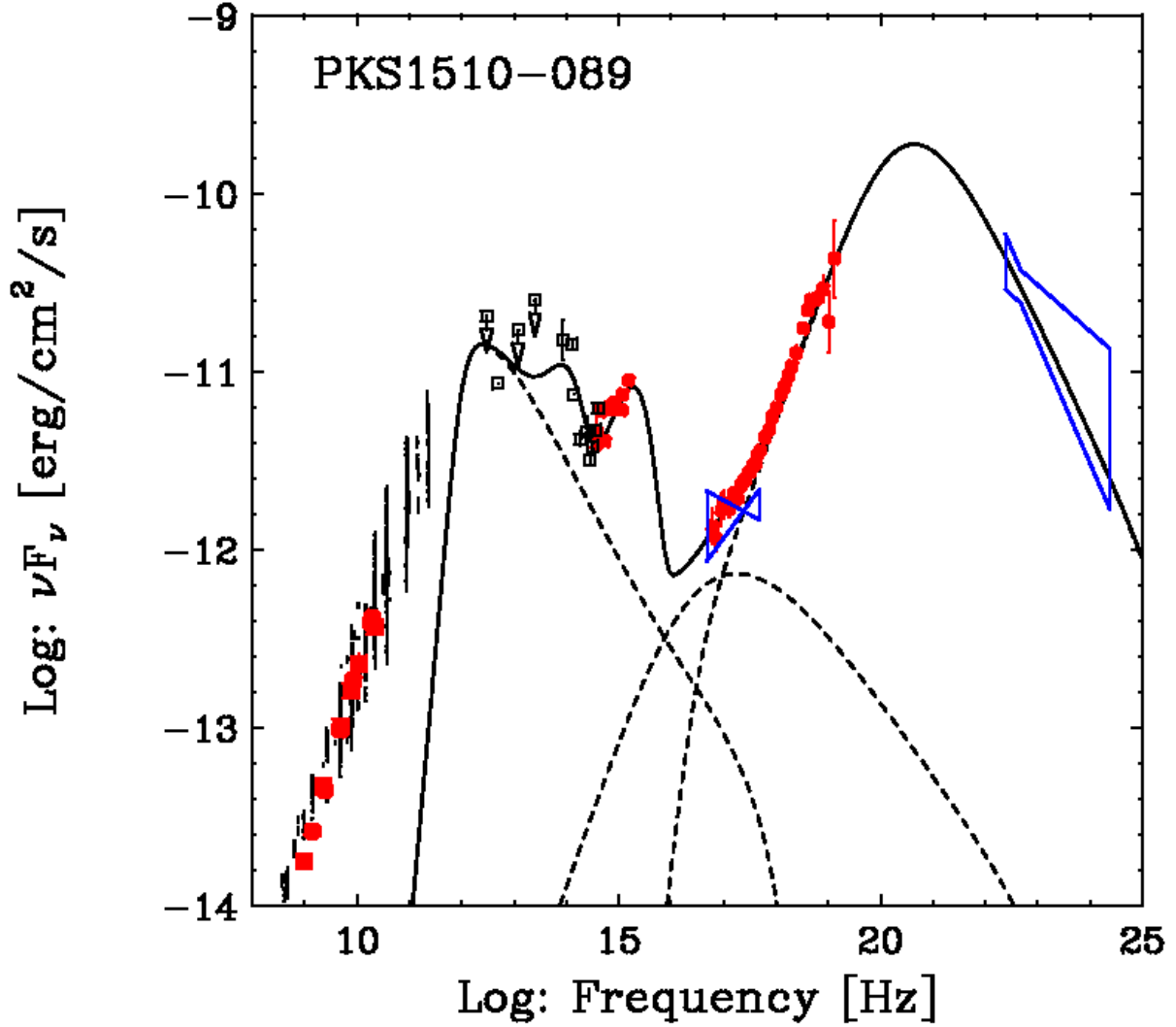}
\caption{Same as Figure 10, but with somewhat different input model parameters:
$\gamma_{\rm min}$ = 1, $\gamma_{\rm br}$ = 150, $\gamma_{\rm max}$
 = 10$^5$, $p$ = 1.35, $q$ = 3.25, $K_e$ = 1.7$\times$10$^{47}$ s$^{-1}$, 
$\Gamma_{\rm jet}$ = 20, $\theta_{\rm jet}$ = 0.02 rd, $\theta_{\rm obs}$ = 
0.05 rd, $r_{\rm sh}$ = 10$^{18}$ cm, $B = 0.86$ G, 
$r_{\rm ext}$ = 3.0$\times$$10^{18}$ cm, $L_{\rm ext}$ = 
3.7$\times$10$^{45}$ erg s$^{-1}$, and $h\nu_{\rm ext}$ 
= 0.2 eV.}\label{fig:MW2}
\end{center}
\end{figure}

\begin{figure}
\begin{center}
\includegraphics[angle=0,scale=.80]{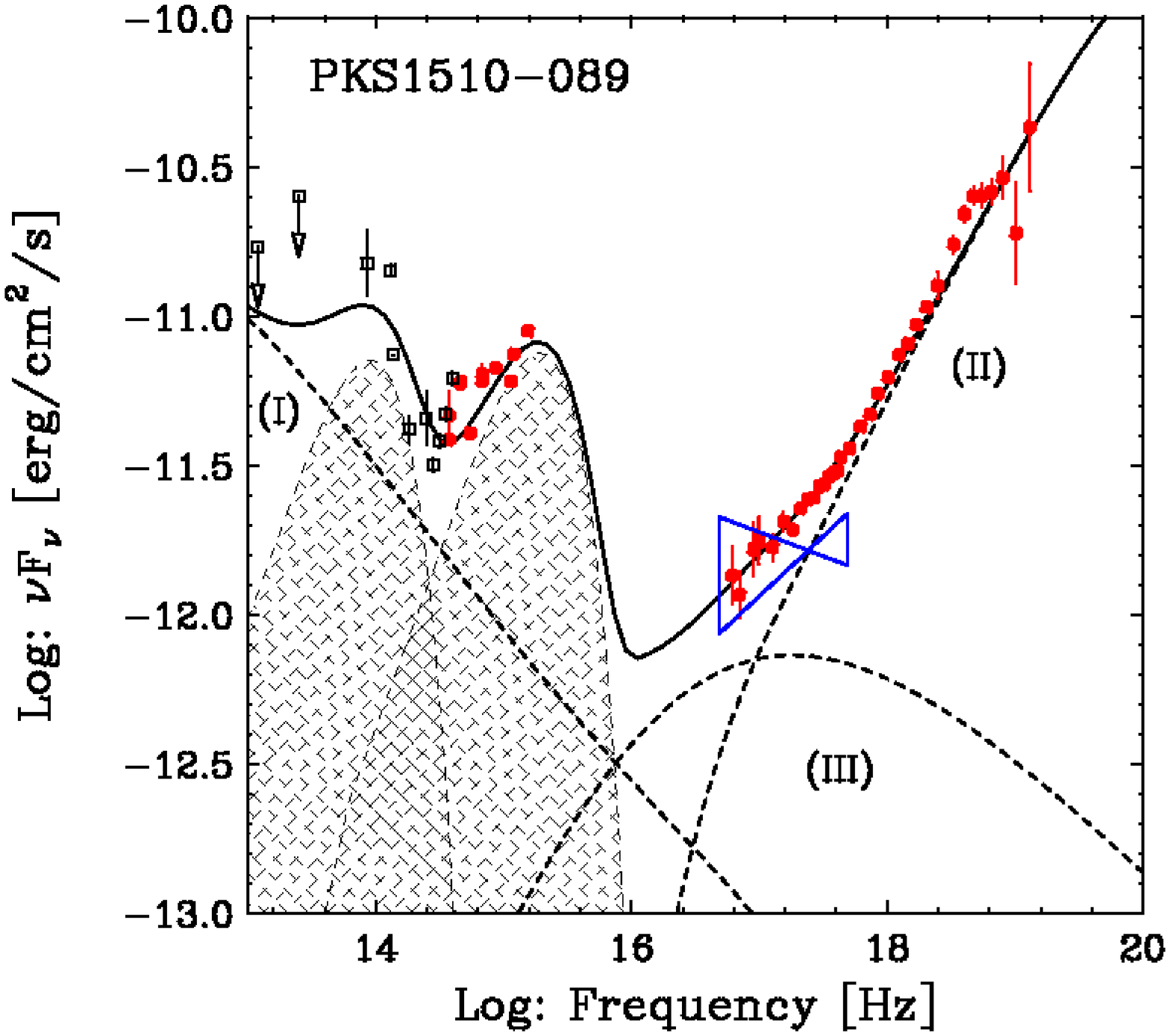}
\caption{A close-up of the PKS~1510$-$089 SED model presented in Figure 12
between the optical and
the X-ray bands. The left hump mimics an excess emission from the 
dusty torus as suggested by {\it IRAS} (Tanner et al. 1996), 
whereas the right bump expected 
for this source from combined {\swift}/UVOT, REM, and Heidelberg data.
Dotted lines show: (I) the Synchrotron; (II) the ERC; and (III) the SSC 
components, respectively. 
A thick line shows sum of all the model components.}\label{fig:MW2_zoom}
\end{center}
\end{figure}


\begin{thebibliography}{}
\bibitem[]{abbey06} Abbey, T. 2006, proceedings of the conference
	  ``The X-ray Universe'', El Escorial, 2005, ESA-SP 604, 943
\bibitem[]{arnaud96} Arnaud, K.~A. 1996, ASP Conf.~Ser.~101:
	  Astronomical Data Analysis Software and Systems V, 101, 17
\bibitem[]{baa77} Baars, J, W. M, Genzel, R., Pauliny-Toth, I. I. K,
 	  Witzel, A. 1977, \aap, 61, 99
\bibitem[]{bar05} Barthelmy, S. D., et al. 2005, Space Sci.\ Rev., 120, 143
\bibitem[]{beg87} Begelman, M. C., \& Sikora, M. 1987, ApJ, 322, 650
\bibitem[]{ber96} Bertin E., \& Arnouts S. 1996, A\&AS, 117, 393
\bibitem[]{bur05} Burrows, D. N., et al. 2005, Space Sci.\ Rev., 120, 165
\bibitem[]{bur82} Burstein, D., \& Heiles, C. 1982, AJ, 87, 1165
\bibitem[]{cap97} Cappi, M., et al. 1997, ApJ, 478, 492
\bibitem[]{car89} Cardelli, J. A., Clayton, G. C., \& Mathis, J. S. 
1989, \apj, 345, 245
\bibitem[]{cel97} Celotti, A., Padovani, P., \& Ghisellini, G. 1997,
	  MNRAS, 286, 415
\bibitem[]{cel07} Celotti, A., Ghisellini, G., \& Fabian, A. C. 2007,
	  MNRAS, 375, 417
\bibitem[]{con04} Conconi, P., et al. 2004, Proc.\ SPIE, 5492, 1602
\bibitem[]{cru06} Crummy, J., Fabian, A. C., Gallo, L., \& Ross, R. R.
	  2006, MNRAS, 365, 1067 
\bibitem[]{der93} Dermer, C. D., \& Schlickeiser, R. 1993, ApJ, 416, 458
\bibitem[]{DN07} Done, C., \& Nayakshin, S. 2007, MNRAS, 377, L59
\bibitem[]{eis97} Eisloffel, J., \& Mundt, R., 1997, AJ, 114, 280
\bibitem[]{Geo05} Georganopoulos, M., Kazanas, D., Perlman, E., \&
	  Stecker, F. 2005, ApJ, 625, 656
\bibitem[]{fuk06} Fukazawa, Y., et al. 2006, Proc.\ SPIE, 6266, 75
\bibitem[]{fra92} Frater, R. H., Brooks, J. W., \& Whiteoak, J. B. 1992, 
                  J.\ Electrical and Electronics Eng.\, Australia, 12, 103
\bibitem[]{fro07} Frontera, F., et al. 2007, ApJ, in press (astro-ph/0611228)
\bibitem[]{gamb03} Gambill, J. K., et al. 2003, A\&A, 401, 505
\bibitem[]{gehrels04} Gehrels, N., et al. 2004, ApJ, 611, 1005
\bibitem[]{ghi98} Ghisellini, G., Celotti, A., Fossati, G., Maraschi,
	  L., \& Comastri, A. 1998, MNRAS, 301, 451
\bibitem[]{gru99} Gruber, D. E., Matteson, J. L., Peterson, L. E., \& 
        Jung, G. V. 1999, ApJ, 520, 124
\bibitem[]{har99} Hartman R.C., et al. 1999, ApJS, 123, 79
\bibitem[]{hil04} Hill, J.E., et al., 2004, Proc.\ SPIE,
5165, 217
\bibitem[]{hir05} Hirotani, K. 2005, ApJ, 619, 73
\bibitem[]{hom01} Homan D. C., et al. 2001, ApJ, 549, 840
\bibitem[]{hos92} Hoshino, A., Arons, J., Gallant, Y. A., \& Langdon,
	  A. B. 1992, ApJ, 390, 454
\bibitem[]{jor05} Jorstad, S. G., et al.  2005, AJ, 130, 1418
\bibitem[]{kat99} Kataoka, J., et al.  1999, ApJ, 514, 138
\bibitem[]{kat01} Kataoka, J., et al.  2001, ApJ, 560, 659
\bibitem[]{kat07} Kataoka, J., et al.  2007, PASJ, 59, 279
\bibitem[]{koi99} Koide, S., Meier, D., Shibata, K., \& Kudoh, T. 1999,
	  ApJ, 536, 668
\bibitem[]{kok07} Kokubun, M., et al. 2007, PASJ, 59, 53
\bibitem[Korolkov \& Pariiskii(1979)]{RATAN} Korolkov, D.~V., \&
	  Pariiskii,  I.~N.\ 1979, \skytel, 57, 324
\bibitem[Kovalev et~al.(1999)]{Kov99}
Kovalev, Y. Y., Nizhelsky, N. A., Kovalev, Yu. A., Berlin, A. B.,
	  Zhekanis, G. V., Mingaliev, M. G., \& Bogdantsov, A. V. 1999,
	  \aaps, 139, 545
\bibitem[]{kom07} Komissarov, S.S., Barkov, M.V., Vlahakis, N., \& 
K\"onigl, A. 2007, MNRAS, in press [arXiv:astro-ph/0703146v2]
\bibitem[]{koy07} Koyama, K., et al. 2007, PASJ, 59, 23
\bibitem[]{kub98} Kubo, H., Takahashi, T., Madejski, G., Tashiro, M.,
	  Makino, F., Inoue, S., \& Takahara, F. 1998, ApJ, 504, 693 
\bibitem[]{laor97} Laor, A., Fiore, F., Elvis, M., et al. 1997, ApJ, 477, 93
\bibitem[]{law92} Lawson, A. J., Turner, M. J. L., Williams, O. R.,
	  Stewart, G. C., \& Saxton, R. D. 1992, MNRAS, 259, 743
\bibitem[]{liu06} Liu, Y., Jiang, D. R., \& Gu, M. F. 2006, ApJ, 637, 669
\bibitem[]{loc95} Lockman, F. J., \& Savage, B. D. 1995, ApJS, 97, 1  
\bibitem[]{lov99} Lovelace, R. V. E., Ustyugova, G. V., \& Koldova , 
A. V. 1999, {\it {Active Galactic Nuclei and Related Phenomena}}, IAU 
Symposium, 194, 208  
\bibitem[]{mad99} Madejski, G., et al. 1999, ApJ, 521, 145
\bibitem[]{mak91} Makino, F., et al. 1989, ApJ, 347, L9  
\bibitem[]{mal86} Malkan, M. A., \& Moore, R. L. 1986, ApJ, 300, 216 
\bibitem[]{mir99} Mirabel, I. F., \& Rodriguez, L. F. 1999, ARA\&A, 37,
	  409
\bibitem[]{mit84} Mitsuda, K., et al. 1984, PASJ, 36, 741
\bibitem[]{mit07} Mitsuda, K., et al. 2007, PASJ, 59, 1
\bibitem[]{mod03} Moderski, R., Sikora, \& B\l a{\.z}ejowski, M. 2003,
	          A\&A, 406, 855
\bibitem[]{ms07} Moderski, R., \& Sikora, M. 2007, to appear in 
                 "The Central Engine of Active Galactic Nuclei", 
                 ed. L. C. Ho and J.-M. Wang (San Francisco: ASP),
                 arXiv:astro-ph/0612342
\bibitem[]{mod04} Moderski, R., Sikora, M., Madejski, G. M., \& Kamae,
	          T. 2004, ApJ, 611, 770
\bibitem[]{mod05} Moderski, R., Sikora, M., Coppi, P. S., \& Aharonian, F. 
                  2005, MNRAS, 363, 954
\bibitem[]{ott94} Ott, M., Witzel, A., Quirrenbach, A., Krichbaum, T. P., 
                  Standke, K. J., Schalinski, C. J., \& Hummel, C. A. 1994,
                  \aap, 284, 331
\bibitem[]{pag05} Page, K. L., Reeves, J. N., O'Brien, P. T., \& Turner,
	  M. J. L. 2005, MNRAS, 364, 195
\bibitem[]{pia93} Pian, E. \& Treves, A. 1993, \apj, 416, 130
\bibitem[]{pir00} Piran, T. 2000, Physics Reports, 333, 529
\bibitem[]{pol04} Porquet, D., Reeves, J. N., O'Brien, P., \& Brinkmann,
	  W., 2004, A\&A, 422, 85
\bibitem[]{rai98} Raiteri, C. M., Villata, M., Lanteri, L., Cavallone,
	  M., \& Sobrito, G. 1998, A\&AS, 130, 495
\bibitem[]{rom92} Romanova, M. M., \& Lovelace, R. V. E. 1992, A\&A,
	  262, 26
\bibitem[]{rom05} Roming, P. W. A., et al.  2005, Space Sci.\ Rev., 120, 95
\bibitem[]{sam06} Sambruna, R., et al. 2006, ApJ, 646, 23 
\bibitem[]{sam06} Sambruna, R., et al. 2007, ApJ, in press	  
\bibitem[]{sau95} Sault, R. J., Teuben, P. J., \& Wright, M. C. H.,
	  1995, Astronomical Data Analysis Software and Systems IV, 77, 433
\bibitem[]{ser07} Serlemitsos, P. J., et al. 2007, PASJ, 59, 9
\bibitem[]{shl98} Schlegel, D. J., Finkbeiner, D. P., \& Davis. M. 1998,
	  ApJ, 500, 525
\bibitem[]{sie96} Siebert, J., et al. 1996, MNRAS, 279, 1331
\bibitem[]{sik94} Sikora, M., Begelman, M. C., \& Rees, M. J., 
1994, \apj, 421, 153
\bibitem[]{sik97} Sikora, M., Madejski, G.  Moderski, R., \& Poutanen,
	  J. 1997, ApJ, 484, 108 
\bibitem[]{sik00} Sikora, M. \& Madejski, G. M. 2000, \apj, 534, 109
\bibitem[]{sik02} Sikora, M., B\l a\.zejowski, M., Moderski, R., \&
	  Madejski, G. M. 2002, ApJ, 577, 78
\bibitem[]{sik05} Sikora, M., Begelman, M. C., Madejski, G. M., \&
	  Lasota, J.-P. 2005, ApJ, 625, 72
\bibitem[]{sin97} Singh, K. P., Shrader, C. R., \& George, I. M. 1997,
	  ApJ, 365, 455
\bibitem[]{sok05} Sokolov, A., \& Marscher, A.P. 2005, ApJ, 629, 52
\bibitem[]{spa00} Spada, M., Panaitescu, A., \& Meszaros, P. 2000, 
ApJ, 537, 824
\bibitem[]{sta07} Stawarz, \L ., Cheung, C. C., Harris, D. E., \& Ostrowski, M.
2007, ApJ, 662, 213
\bibitem[]{ste88} Stetson P. B. 1988, PASP, 99, 191
\bibitem[]{tag00} Tagliaferri, G., et al. 2000, A\&A, 354, 431
\bibitem[]{tak07} Takahashi, T., et al. 2007, PASJ, 59, 35
\bibitem[]{tan00} Tanihata, C., et al. 2000, ApJ, 543, 124
\bibitem[]{tan03} Tanihata, C., Takahashi, T., Kataoka, J., \& Madejski,
	  G. M. 2003, ApJ, 584, 153
\bibitem[]{tav00} Tavecchio, F., et al. 2000, \apj, 543, 535
\bibitem[]{tav07} Tavecchio, F., Maraschi, L., Ghisellini, G., Kataoka,
	  J., Foschini, L., Sambruna, R. M., \& Tagliaferri, G. 2007,
	  \apj, 665, 980
\bibitem[]{tan96} Tanner, A. M., Bechtold, J., Walker, C. E., Black, J. H., 
                  \& Cutri, R. M. 1996, AJ, 112, 62
\bibitem[]{tin03} Tingay, S. J., Jauncey, D. L., King, E. A.,
	  Tzioumis, A. K., Lovell, J. E. J., \& Edwards, P.~G. 2003,
	  \pasj, 55, 351
\bibitem[]{tos96} Tosti, G., Pascolini, S., \& Fiorucci, M. 1996, PASP,
	  108, 706
\bibitem[]{tos04} Tosti, G., et al. 2004, Proc.\ SPIE, 5492, 689
\bibitem[]{uch06} Uchiyama, Y., et al. 2005, ApJ, 631, L113 
\bibitem[]{urr95} Urry, C. M., \& Padovani, P. 1995, PASP, 107, 803
\bibitem[]{ver97} Verkhodanov, O. V., Trushkin, S. A., Andernach, H., 
                  \& Chernenkov, V. N. 1997, Astronomical Data Analysis 
                  Software and Systems VI, 125, 322
\bibitem[]{vil97} Villata M., et al. 1997, A\&AS, 121, 119
\bibitem[]{wal93} Walter, R., \& Fink, H. H. 1993, A\&A, 274, 105
\bibitem[]{war98} Wardle, J. F. C., Homan, D. C., Ojha, R., \& Roberts, 
D. H. 1998, Nature, 395, 457 
\bibitem[]{war02} Wardle, J. F. C., Homan, D. C., Cheung, C. C.,
	  Roberts, D. H. 2005, in ASP Conf.\ Ser.\ 340, 
          Future Directions in High Resolution Astronomy: The 10th Anniversary of the VLBA, ed. J. Romney, \& M. Reid (San Frqancisco, ASP) 67 
\bibitem[]{wil92} Williams, O. R., et al. 1992, ApJ, 389, 157
\bibitem[]{zer04} Zerbi, F. M., et al. 2004,  Proc.\ SPIE, 5492, 1590 
\bibitem[]{zdz00} Zdziarski, A. A., Poutanen, J., \& Johnson, W. N. 
                  2000, \apj, 542, 703
\end{thebibliography}
\end{document}